\documentclass[a4paper,12pt]{article}

\usepackage{amsmath}
\usepackage{amssymb}
\usepackage{amsthm}

\newcommand{\prt}{\partial}

\def\Pr{{\mathbb P}}
\def\F{{\mathcal F}}

\def\ni{{\noindent}}

\def\be{\begin{equation}}
\def\ee{\end{equation}}

\theoremstyle{definition}

\begin{document}
\bibliographystyle{plain}

\title{{\Large\bf  Beta ensembles, quantum Painlev\'e equations and isomonodromy systems.}}
\author{Igor Rumanov \\  %\footnote{e-mail: igor.rumanov@colorado.edu}} 
{\small Dept. of Applied Mathematics, CU Boulder, Boulder, CO} \\
{\small e-mail: igor.rumanov@colorado.edu} }

\maketitle

\bigskip

\begin{abstract}
This is a review of recent developments in the theory of beta ensembles of random matrices and their relations with conformal filed theory (CFT). There are (almost) no new results here. %We describe the context in which quantum Painlev\'e and more general multidimensional linear equations of Belavin-Polyakov-Zamolodchikov (BPZ) type occupy the main stage. 
This article can serve as a guide on appearances and studies of quantum Painlev\'e and more general multidimensional linear equations of Belavin-Polyakov-Zamolodchikov (BPZ) type in literature. We demonstrate how BPZ equations of CFT arise from $\beta$-ensemble eigenvalue integrals. Quantum Painlev\'e equations are relatively simple instances of BPZ or confluent BPZ equations, they are PDEs in two independent variables (``time" and ``space"). While CFT is known as quantum integrable theory, here we focus on the appearing links of $\beta$-ensembles and CFT with {\it classical} integrable structure and isomonodromy systems. The central point is to show on the example of quantum Painlev\'e II (QPII)~\cite{betaFP1} how classical integrable structure can be extended to general values of $\beta$ (or CFT central charge $c$), beyond the special cases $\beta=2$ ($c=1$) and $c\to\infty$ where its appearance is well-established. We also discuss an \'a priori very different important approach, the ODE/IM correspondence giving information about complex quantum integrable models, e.g.~CFT, from some stationary Schr\"odinger ODEs. Solution of the ODEs depends on (discrete) symmetries leading to functional equations for Stokes multipliers equivalent to discrete integrable Hirota-type equations. The separation of ``time" and ``space" variables, a consequence of our integrable structure, also leads to Schr\"odinger ODEs and thus may have a connection with ODE/IM methods. 
\end{abstract}

\newpage

\section{Introduction}

Beta ensembles of random matrices (RM), introduced by Dyson~\cite{DyBeta} in 1962, find more and more applications in physics and mathematics, e.g.~in conformal and integrable quantum field theories (QFT), theory of second order phase transitions, condensed matter theory in connection with conduction in disordered wires, Quantum Hall effect, anyons and fractional (exclusion) statistics. A rather comprehensive treatment can be found in~\cite{F2010}. In fact, their wide applicability originates in their very definition~\cite{DyBeta} as Coulomb gas (fluid) of particles-eigenvalues on the real line or in complex plane. Here we would like to concentrate on their properties related to integrability and integrable systems -- both classical and quantum, with the aim toward the exact solvability of the related problems, which partly explains the term integrability here -- it is first and foremost exact solvability for us. What we mean by exact solvability should become clearer below in section 4.
\par There are two kinds of integrable systems (IS), often considered separately and studied by quite different methods -- classical and quantum IS. The former usually involve functions and differential equations they satisfy while the latter usually deal with non-commutative algebras of operators, their eigenvalues and eigenfunctions, where the main tools are often representation-theoretic. The common integrable structure, however, is clearly seen in the existence of Lax matrices and compatibility conditions. Albeit, while classical IS usually deal with number-valued Lax matrices, for quantum IS the Lax matrices are operator-valued. A general paradigm coming from physics is to consider a quatum IS as quantization of a classical IS. While arisen historically to become a common lore, quantization procedure is often not well defined mathematically and ambiguous due to operator ordering issues. Integrability cures many of these issues, and quantum integrable theories are often better defined than other quantum theories (this is especially true of QFT). 
\par The relationship between quantum and classical IS, however, goes deeper than just the latter being a classical limit of the former. An exact, without any limit, correspondence between quantum transfer matrices and classical discrete Hirota equations was first found in~\cite{KLWZ97}, and later in many other cases, see e.g.~\cite{TYsys} and references therein. In another related venue, the ODE/IM (ordinary differential equations/integrable (quantum) models) correspondence was revealed in~\cite{DorTat99, BLZ01, BLZ03}, see~\cite{ODEIM} for a review. There the solution of Baxter equations for quantum transfer matrices of quantum spin chains, or transfer operators in CFT~\cite{BLZ}, and their eigenvalues hinged upon finding energy spectra of certain stationary Schr\"odinger equations (SE). This led to the algebraic Bethe equations for the energy eigenvalues and again to discrete Hirota equations for the Stokes multipliers identified with the eigenvalues of quantum transfer matrices. We are aiming to demonstrate that Dyson beta ensembles and related canonically quantized Painlev\'e or Garnier equations are excellent natural models for further insight into these deep connections. In this review, we describe important related results focusing on the relatively simple examples.
\par Plan of the paper is as follows. Section 2 introduces the main objects of study. Simple derivation of BPZ-type equations for $\beta$-ensembles is given in section 3. In the central section 4, several representative examples of exact integrability in conformal field theory (CFT) and $\beta$-ensemble problems for general Dyson index $\beta$ are outlined. The main conjecture based on the results of~\cite{betaFP1} is that this integrability can be generalized to achieve everything mentioned in section 5 considering special cases $\beta=2$, $\beta=0$ and $\beta\to\infty$ where classical integrability is well-known. Section 6 briefly touches on the appearance of more general PDEs involving multidimensional diffusion-drift operators, e.g.~of quantized Garnier type. Section 7 contains some concluding remarks.

\subsection{Work of Dyson and further developments}

Dyson considered~\cite{DyBeta} the Brownian Motion (BM) of eigenvalues of Gaussian $\beta$-ensemble defined by eigenvalue distribution
%matrix entries~\cite{DyBeta}, $M_i \equiv M_{jk}$, where $i = (n-1)j + k$ is in the range $i = 1, \dots, N = n^2$, of the matrix $M$, with Gaussian invariant measure

%$$
%\Pr(M_1, \dots, M_N) = c_0\exp\left(-\frac{\beta\text{Tr}M^2}{2a^2}\right),
%$$

%\ni with $c_0$ -- a numerical constant. Then it is easy to see that the average

%$$
%\langle M_i^2\rangle = \frac{a^2}{2\beta}g_i, \ \ \ g_i = 1 + \delta_{jk}.
%$$

%\ni with a normalization constant $c_0$. The corresponding eigenvalue distribution is given by

$$
\Pr(x_1, \dots, x_n) \sim |\Delta|^{\beta}\exp\left(-\frac{\beta\sum_ix_i^2}{2a^2}\right) = e^{-\beta W},   \eqno(1.1)
$$

\ni where the potential $W$ is

$$
W(x_1,\dots, x_n) = -\sum_{i<j}\ln|x_i - x_j| + \sum_i\frac{x_i^2}{2a^2}.
$$

\ni This is equivalent to the dynamics of particles in Brownian motion with positions $x_i$, subjected to an electric force $E(x_i)=-\prt_{x_i}W$ and friction with strength $f$, such that at temperature $T$ during a small time interval $\delta t$ changes in particle positions $\delta x_i$ are given by

%$$
%E(x_i) = -\frac{\prt W}{\prt x_i} = \sum_{i\ne j}\frac{1}{x_i - x_j} - \frac{x_i}{a^2},
%$$

%\ni and friction with strength $f$, such that during a small time interval $\delta t$ changes in particle positions $\delta x_i$ are given by

$$
f\langle\delta x_i\rangle = E(x_i)\delta t, \ \ \  f\langle\delta x_i^2\rangle = 2T\delta t,
$$

\ni and all higher moments are zero. The joint probability density (p.d.f.) $\Pr(x_1, \dots, x_n; t)$ then satisfies the Fokker-Planck (FP) equation %with diffusion and drift terms,

$$
f\frac{\prt \Pr}{\prt t} = \sum_i \left(T\frac{\prt^2\Pr}{\prt x_i^2} - \frac{\prt}{\prt x_i}(E(x_i)\Pr)\right).
$$

\ni Its unique stationary solution is (1.1), i.e.~the original random matrix (RM) eigenvalue distribution. The respective Brownian motinon of matrix elements is defined for $\beta=1, 2$ or $4$ only\footnote{Nowadays, however, starting with the seminal work~\cite{DE02}, many matrix models leading to general $\beta$ eigenvalue distributions are known~\cite{ABG12, EdEtAlBeta13, KRV13}.} and satisfies another FP equation,
%determined by %corresponding to this particle system, 

%$$
%f\langle\delta M_i\rangle = -\frac{M_i}{a^2}\delta t, \ \ \  f\langle\delta M_i^2\rangle = g_i\cdot T\delta t,
%$$

%\ni which implies another FP equation for the matrix entries p.d.f. $\Pr(M_1, \dots, M_N; t)$,

$$
f\frac{\prt \Pr}{\prt t} = \sum_i \left(\frac{T}{2}g_i\frac{\prt^2\Pr}{\prt M_i^2} + \frac{1}{a^2}\frac{\prt}{\prt M_i}(M_i\Pr)\right).
$$

\ni For an initial condition as $M = M'$ at $t = 0$, the unique solution of the last equation is

$$
\Pr(M; t) = \frac{c_0}{(1 - c^2)^{N/2}}\exp{\left(-\frac{\text{Tr}(M - cM')^2}{2a^2k_BT(1-c^2)} \right)},
$$

\ni where $c_0$ is a numerical constant and $c = e^{-\frac{t}{a^2f}}$. I.e.~here $\beta=1/T$. The Brownian Motion (BM) is invariant under symmetry-preserving unitary transformations of $M$. Dyson showed that the two BMs correspond to each other in the sense that the first is the BM of eigenvalues for the matrix whose entries are involved in the second one. In applications, e.g.~in studies of conductivity of wires with impurities creating disorder, a similar model often arises, that of a random matrix $H$ consisting of random and non-random components, %JohBM, 

$$
H = |1-e^{-2\tau}|^{1/2}H_{rand.} + e^{-\tau}H_0,
$$

\ni where $H_0$ is non-random, $\tau$ is the strength of disorder playing the role of time for Dyson BM~\cite{SLAl93, F2010}. %Its eigenvalue p.d.f.~also satisfies a FP equation, see e.g.~\cite{F2010},

%$$
%\prt_\tau P(\tau, x_1, \dots, x_N) = \left( \frac{1}{\beta}\sum_{k=1}^N\prt_{x_k}^2 + \sum_{k=1}^N\prt_{x_k}\left(x_k - \sum_{j\neq k}^N\frac{1}{x_k-x_j}\right) \right)P(\tau, x_1, \dots, x_N).
%$$

%General probabilistic argument for going from 1st order SDE to FP equation:

\ni These FP equations are autonomous and therefore admit simple solutions. It turns out that the whole related nonlinear integrable structure can be derived from this type FP equations but {\it nonautonomous} ones~\cite{ZZFP, betaFP1}, which is one of the main subjects of the present paper. Besides, one is more often interested in the integrals of the p.d.f. rather than the p.d.f. itself. The integrals sometimes also satisfy certain FP equations which we consider in section 3.
%\par {\bf Physics:} disordered wires, (fractional) quantum Hall effect, QFT (gauge 4d SUSY QFT $\to$ 2d CFT), 2nd order phase transitions in 2 dim. : CFT, SLE, (circular) Dyson BM
%\par ...

\section{Beta ensembles, Virasoro constraints and quantum integrable models}

Consider the integrals defining $\beta$-ensembles,

$$
I_\beta([t]) = \int\dots\int \prod_{i=1}^Mdx_i |\Delta(x)|^\beta e^{\sum_{k=0}^\infty t_k\sum_{i=1}^Mx_i^k}, \qquad \Delta(x) = \prod_{i<j}(x_i-x_j).   \eqno(2.1)
$$

\ni The integration here can be considered over (subsets of) $\mathbb R^M$. The integral satisfies the so-called Virasoro constraints derived for general $\beta$ in~\cite{AwMOSh}, exposing the connection of $\beta$-ensembles and conformal filed theory (CFT) with central charge

$$
c = 1 - 6\frac{(1-\kappa)^2}{\kappa}.   \eqno(2.2)   %(2.3)
$$

\ni The following identity expresses the Virasoro constarints 

$$
0 = \int\dots\int \prod_{i=1}^Mdx_i \sum_{i=1}^M\frac{\prt}{\prt x_i}\left(x_i^{n+1}|\Delta(x)|^\beta e^{\sum_{k=0}^\infty t_k\sum_{i=1}^Mx_i^k}\right) = L_nI_\beta([t]), \quad (n=-1, 0, 1, \dots),   \eqno(2.3)   %(2.4)
$$

\ni where

$$
L_n = \kappa\sum_{m=0}^n\frac{\prt^2}{\prt t_m\prt t_{n-m}} + \sum_{m=1}^\infty mt_m\frac{\prt}{\prt t_{n+m}} + (1-\kappa)(n+1)\frac{\prt}{\prt t_n}   \eqno(2.4)  %(2.5)
$$

\ni are the Virasoro generators making an infinite subalgebra ($m, n \ge -1$) of the Virasoro algebra with commutation relations

$$
[L_n, L_m] = (n-m)L_{n+m} + \frac{c}{12}n(n^2-1)\delta_{n,-m}.  \eqno(2.5)  %(2.6)
$$

\par The connection of integral (2.1) with CFT is best demonstrated by introducing so-called free collective field operators, or holomorphic ({\it chiral}) bosons~\cite{AwMOSh, Kostov99}:

$$
\prt\phi(z) = \sqrt\beta \sum_{n=0}^\infty z^{-n-1}\frac{\prt}{\prt t_n} + \frac{1}{\sqrt\beta}\sum_{n=1}^\infty nt_nz^{n-1}.   \eqno(2.6)  %(2.7)
$$

\ni Then the generators $L_n$ in (2.4) are the Laurent (or Fourier) modes of the holomorphic component of the CFT energy-momentum tensor~\cite{BPZ, DF, Kostov99}

$$
T(z) = \sum_{-\infty}^\infty \frac{L_n}{z^{n+2}} = \frac{1}{2} :(\prt\phi(z))^2: - \frac{1-\kappa}{\sqrt{2\kappa}}\prt^2\phi(z),  \eqno(2.7)  %(2.8)
$$

\ni where here and further on $\kappa=\beta/2$, colons denote the necessary normal ordering (i.e.~putting all the $t_n$ derivatives to the right) to make square of the operator generating series (2.6) well-defined.
\par There is also a direct relation of certain integrals of the form (2.1) with quantum Calogero-Sutherland model (CSM)~\cite{F2010, AwMOSh} defined by Hamiltonian for $N$ particles,

$$
\mathcal H = -\frac{1}{2}\sum_{k=1}^N\frac{\prt^2}{\prt q_k^2} + \frac{\pi^2}{2L^2}\sum_{j\neq k}^N\frac{\kappa(\kappa-1)}{\sin^2 \pi(q_j-q_k)/L},  \eqno(2.8)  %(2.2)
$$

\ni Following~\cite{AwMOSh, F2010}, denote $y_j=e^{2\pi iq_j/L}$ and consider another multidimensional integral, the Selberg-Aomoto integral,

$$
S_{M,N}(a,b,\gamma,\mu; [y]) = \int_{[0,1]^M} \prod_{i=1}^Mdx_i \cdot \prod_{i=1}^M\prod_{k=1}^N(1-x_iy_k)^\alpha \prod_{i=1}^Mx_i^a(1-x_i)^b\prod_{i<j}^M|x_i-x_j|^\gamma,  \eqno(2.9)
$$

\ni which is essentially an averaged power $\alpha$ of the product of $N$ characteristic polynomials of size $M$ Jacobi $\beta$-ensemble. The correspondence between (2.9) and (2.1) is established via the transformation of symmetric variables $y_j$ into the power sums $t_k=\sum_{j=1}^Ny_j^k/k$,

$$
\prod_{i=1}^M\prod_{n=1}^N(1-x_iy_n)^\alpha = \prod_{i=1}^M e^{-\alpha\sum_{k=1}^\infty t_kx_i^k}.   \eqno(2.10)
$$

\ni Thus, for special points in the infinite space of couplings $\{t_k\}$, such that only the first $N$ of them are independent, the integral (2.1) reduces to the one of the form (2.9) or (3.1) below.
\par On the other hand, the CSM Hamiltonian (2.8) can be transformed as

$$
\tilde\Delta(y)^{-\kappa}\mathcal H\tilde\Delta(y)^{\kappa} = 2\left(\frac{\pi}{L}\right)^2\tilde H + E_0, \quad \tilde\Delta(y) = \prod_{i<j}^N\sin \pi(q_i-q_j)/L \sim \prod_iy_i^{-(N-1)/2}\prod_{i<j}^N(y_i-y_j),   \eqno(2.11)
$$

\ni where

$$
\tilde H = \sum_{i=1}^N(y_i\prt_{y_i})^2 + \kappa\sum_{i<j}^N\frac{y_i+y_j}{y_i-y_j}(y_i\prt_{y_i} - y_j\prt_{y_j}),   \eqno(2.12)
$$

\ni and $E_0=(\pi/L)^2\kappa^2(N^3-N)/6$ is the eigenvalue of the ground state $\tilde\Delta(y)^{\kappa}$. The eigenfunctions of $\tilde H$ are Jack polynomials, see e.g.~\cite{Mac, F2010}. The integral (2.9) can be expanded in a (convergent) infinite linear combination of Jack polynomials in $y$-variables with $\kappa=2\alpha^2/\gamma$, and so is an eigenfunction of the operator $\tilde H$. It satisfies a multivariate generalization of hypergeometric differential equation when $\alpha = 1$ or $\alpha = -\gamma/2$, see e.g.~\cite{F2010}, a special case of PDEs considered in the next section. %Then the integral (2.1) is a more general eigenfunction of $\tilde H$ and also a certain correlation function of the quantum CSM theory (???).
%\par About cut-and-join operators and single Hurwitz generating functions ...

\section{BPZ differential equations for $\beta$-ensembles and CFT~\cite{MMM11, MarTak, AwEtAl10, AgEtAl11, Nag11}}

As an example for PDE derivation, consider an integral

%$$
%Z = e^{\gamma V(z)}\int\dots\int \prod_{i=1}^Ndx_i (z-x_i)^{\alpha}\Delta^{\beta}(x)e^{-\sum_{k=1}^NV(x_k)}.   \eqno(Zint)
%$$

$$
Z = \int\dots\int \prod_{i=1}^Ndx_i (z-x_i)^{\alpha}\Delta^{\beta}(x)e^{-\sum_{k=1}^NV(x_k)}.   \eqno(3.1)  %(Zint)   %\gamma=0
$$

\ni Denoting by $\langle\ \rangle$ the integration with the measure defined by (3.1) and differentiationg $Z$ (so that e.g.~$Z = \langle \rangle$), one has

%$$
%\prt_zZ = \gamma V'(z)Z + \left\langle \sum_{k=1}^N\frac{\alpha}{z-x_k} \right\rangle, \eqno(dZ)
%$$

$$
\prt_zZ = \left\langle \sum_{k=1}^N\frac{\alpha}{z-x_k} \right\rangle, \eqno(3.2)  %(dZ)
$$

%$$
%\prt_{zz}Z = (\gamma V''(z) + \gamma^2(V'(z))^2)Z + 2\gamma V'(z)\prt_zZ + \left\langle \sum_{k=1}^N\left(\frac{\alpha^2-\alpha}{(z-x_k)^2} + \sum_{j\neq k}^N \frac{2\alpha^2}{(z-x_k)(x_k-x_j)}\right) \right\rangle.  \eqno(d2Z)  %,
%$$

$$
\prt_{zz}Z = \left\langle \sum_{k=1}^N\left(\frac{\alpha^2-\alpha}{(z-x_k)^2} + \sum_{j\neq k}^N \frac{2\alpha^2}{(z-x_k)(x_k-x_j)}\right) \right\rangle.  \eqno(3.3) %(d2Z)  %,
$$

\ni On the other hand, consider the identity (it can be considered as a generating function of the Virasoro constraints (2.3), (2.4))

$$
0 = \sum_{k=1}^N \int\dots\int \prod_{i=1}^Ndx_i \frac{\prt}{\prt x_k}\left(\frac{1}{z-x_k}\prod_{i=1}^N(z-x_i)^{\alpha}\Delta^{\beta}(x)e^{-\sum_{k=1}^NV(x_k)}\right) =
$$

$$
= \left\langle \sum_{k=1}^N\left(\frac{1-\alpha}{(z-x_k)^2} + \sum_{j\neq k}^N \frac{\beta}{(z-x_k)(x_k-x_j)} - \frac{V'(x_k)}{z-x_k} \right)\right\rangle.   \eqno(3.4) %(VirZ)
$$

\ni Comparing (3.4) with (3.3), one finds another identity (by subtracting $\alpha\cdot$(3.4) from (3.3)):

%$$
%(\prt_{zz} + 2\gamma V'(z)\prt_z + \gamma V''(z) + \gamma^2(V'(z))^2)\langle\ \rangle - \left\langle\sum_{k=1}^N\sum_{j\neq k}^N \frac{\alpha(\beta+2\alpha)}{(z-x_k)(x_k-x_j)}\right\rangle + \left\langle\sum_{k=1}^N\frac{\alpha V'(x_k)}{z-x_k}\right\rangle = 0.   \eqno(Id1)
%$$

$$
\prt_{zz}\langle\ \rangle - \left\langle\sum_{k=1}^N\sum_{j\neq k}^N \frac{\alpha(\beta+2\alpha)}{(z-x_k)(x_k-x_j)}\right\rangle + \left\langle\sum_{k=1}^N\frac{\alpha V'(x_k)}{z-x_k}\right\rangle = 0.   \eqno(3.5)  %(Id1)
$$

\ni Besides, taking another linear combination, $\beta\cdot$(3.2)$+2\alpha^2\cdot$(3.4) leads to one more identity,

%$$
%(\beta\prt_{zz} + \beta(2\gamma V'(z)\prt_z + \gamma V''(z) + \gamma^2(V'(z))^2))\langle\ \rangle + \left\langle\sum_{k=1}^N\frac{\alpha(1-\alpha)(\beta+2\alpha)}{(z-x_k)^2}\right\rangle - \left\langle\sum_{k=1}^N\frac{2\alpha^2 V'(x_k)}{z-x_k}\right\rangle = 0.   \eqno(Id2)
%$$

$$
\beta\prt_{zz}\langle\ \rangle + \left\langle\sum_{k=1}^N\frac{\alpha(1-\alpha)(\beta+2\alpha)}{(z-x_k)^2}\right\rangle - \left\langle\sum_{k=1}^N\frac{2\alpha^2 V'(x_k)}{z-x_k}\right\rangle = 0.   \eqno(3.6)  %(Id2)
$$

\ni The identities (3.5) and (3.6) can be further simplified by taking the appropriate special values of $\alpha$ and killing the complicated averaged sums. E.g.~taking $\alpha=1$ in (3.6) and using (3.2) yields

%$$
%\left(\frac{\beta}{2}\prt_{zz} + \frac{\beta}{2}(2\gamma V'(z)\prt_z + \gamma V''(z) + \gamma^2(V'(z))^2))\right)\langle\ \rangle - V'(z)(\prt_z-\gamma V'(z))\langle\ \rangle + \left\langle\sum_{k=1}^N\frac{V'(z) - V'(x_k)}{z-x_k}\right\rangle = 0.   \eqno(1Id)
%$$

$$
\frac{\beta}{2}\prt_{zz}\langle\ \rangle - V'(z)\prt_z\langle\ \rangle + \left\langle\sum_{k=1}^N\frac{V'(z) - V'(x_k)}{z-x_k}\right\rangle = 0.   \eqno(3.7)  %(1Id)
$$

\ni Similarly, taking $\alpha=-\beta/2$ in (3.5) leads to

%$$
%(\prt_{zz} + 2\gamma V'(z)\prt_z + \gamma V''(z) + \gamma^2(V'(z))^2)\langle\ \rangle + V'(z)(\prt_z-\gamma V'(z))\langle\ \rangle + \frac{\beta}{2}\left\langle\sum_{k=1}^N\frac{V'(z) - V'(x_k)}{z-x_k}\right\rangle = 0.   \eqno(2Id)
%$$

$$
\prt_{zz}\langle\ \rangle + V'(z)\prt_z\langle\ \rangle + \frac{\beta}{2}\left\langle\sum_{k=1}^N\frac{V'(z) - V'(x_k)}{z-x_k}\right\rangle = 0.   \eqno(3.8)  %(2Id)
$$

\ni The last term in (3.7) or (3.8) for many important potentials can be written explicitly in terms of derivatives of $Z$ w.r.t.~the coupling parameters. For instance, the so-called multi-Penner potential,

$$
V(x) = C(N,\beta)\cdot\sum_{l=1}^{n}m_l\ln(x-w_l).   \eqno(3.9)  %(mPpot)
$$

\ni gives

$$
\left\langle\sum_{k=1}^N\frac{V'(z) - V'(x_k)}{z-x_k}\right\rangle = C(N,\beta)\sum_{l=1}^{n}\frac{m_l}{z-w_l}\left\langle\sum_{k=1}^N\frac{1}{x_k-w_l}\right\rangle = \sum_{l=1}^{n}\frac{1}{z-w_l}\frac{\prt}{\prt w_l}\langle\ \rangle,   \eqno(3.10)  %(t-der)
$$

\ni where we used that $\prt_{w_l}e^{-C\sum_{k=1}^Nm_l\ln(x_k-w_l)} = Cm_l\sum_{k=1}^N1/(x_k-w_l)e^{-C\sum_{k=1}^Nm_l\ln(x_k-w_l)}$. Thus, we obtain the nonlinear PDEs satisfied by $Z$ for the two special values of $\alpha$, $\alpha=1$ and $\alpha=-\beta/2$, respectively:

%$$
%\left(\frac{\beta}{2}\prt_{zz} + (\beta\gamma-1)V'(z)\prt_z + \beta\gamma V''(z)/2 + \gamma(\gamma+\beta/2)(V'(z))^2)) + \sum_{l=0}^{n-2}\frac{m_l}{z-w_l}\frac{\prt}{\prt w_l} \right)\langle\ \rangle_1 = 0,   \eqno(*1)
%$$

$$
\left(\frac{\beta}{2}\prt_{zz} - C(N,\beta)\cdot\sum_{l=1}^{n}\frac{m_l}{z-w_l}\prt_z + \sum_{l=1}^{n}\frac{1}{z-w_l}\frac{\prt}{\prt w_l} \right)\langle\ \rangle_1 = 0,   \eqno(3.11)  %(*1)
$$

%$$
%\left(\frac{2}{\beta}\prt_{zz} + \frac{2}{\beta}((2\gamma+1)V'(z)\prt_z + \gamma V''(z) + \gamma(\gamma-1)(V'(z))^2) + \sum_{l=0}^{n-2}\frac{m_l}{z-w_l}\frac{\prt}{\prt w_l}\right)\langle\ \rangle_{-\beta/2} = 0.   \eqno(*2)
%$$

$$
\left(\frac{2}{\beta}\prt_{zz} + \frac{2}{\beta}C(N,\beta)\cdot\sum_{l=1}^{n}\frac{m_l}{z-w_l}\prt_z + \sum_{l=1}^{n}\frac{1}{z-w_l}\frac{\prt}{\prt w_l}\right)\langle\ \rangle_{-\beta/2} = 0.   \eqno(3.12)  %(*2)
$$

\ni These Fuchsian PDEs are essentially (up to a change due to multiplication of $Z$ (3.1) by a simple factor, see (6.1)) Belavin-Polyakov-Zamolodchikov (BPZ) equations of CFT~\cite{BPZ} corresponding to the insertion of two types of {\it degenerate primary fields} (degenerate primary vertex operators) into a CFT correlation function of a product of $n+1$ non-degenerate primaries (the $\beta$-ensemble integral with $n$ general $w_l$ corresponds to $n$ CFT primaries at their positions and another one implied to be at $\infty$). Our derivation followed~\cite{AgEtAl11, MMM11}, and a similar one was used in~\cite{Nag11} to obtain the connection between all quantum Painlev\'e equations and $\beta$-ensembles with certain potentials. Another important case is that of a polynomial potential $V$, $V(x) = \sum_{l=1}^nt_lx^l$. Then 

$$
\left\langle\sum_{k=1}^N\frac{V'(z) - V'(x_k)}{z-x_k}\right\rangle = \sum_{l=1}^nlt_l\left\langle\sum_{k=1}^Nx_k^{l-1}\right\rangle = \sum_{l=1}^nlt_l\frac{\prt}{\prt t_{l-1}}\langle\ \rangle,   \eqno(3.13)  %(pt-der)
$$

\ni which leads to the {\it confluent} BPZ equations related to {\it irregular} conformal blocks~\cite{JimNagSu08} now realized to be important in asymptotically free gauge theories~\cite{Gai}.
\par Taking $n=3$ and considering two of the $w_l$ as fixed (usually at $0$ and $1$) in (3.11) and (3.12) leads to two quantum Painlev\'e VI equations with two different dual values for the Planck constant, $\hbar=\beta/2$ and $\hbar=2/\beta$, respectively. The connection of the first of them with $\beta$-ensembles was demonstrated e.g.~in~\cite{Nag11}. This explains why the different dual values appeared in different contexts: the $\beta$-ensembles of Nagoya~\cite{Nag11} contain the averaged characteristic polynomials, i.e.~there $\alpha=1$, which leads to PDEs of type (3.11) with $\hbar=\beta/2$. On the contrary, the FP equations derived in the various large $N$ limits of $\beta$-ensembles by probabilists~\cite{EdSut, RRV, RR08, RRZ, BV1, BV2} were related to the ensembles with external matrix source (or with ``spikes" in statistics terminology). Here the formula obtained by Forrester~\cite{ForBeta11} becomes very illuminating. He found the formula for the probability density of finite $N$ Laguerre (Wishart) $\beta$-ensemble with one spike, i.e.~only one eigenvalue of the external source matrix different from $1$,
%formula for the spiked general $\beta$ finite $n$ Wishart matrix p.d.f. was obtained by Forrester~\cite{ForBeta11} based on several earlier results for special cases: 

$$
P_{\beta, a, \delta}(l_1, \dots, l_n) \sim \prod_{i<j}|l_i-l_j|^{\beta}\cdot\prod_{k=1}^nl_k^{(a+1)\beta/2 - 1}e^{-\beta l_k/2}\int_{-\infty}^{+\infty}e^{it}\prod_{k=1}^n\left(it - \frac{\delta-1}{2\delta}l_k\right)^{-\beta/2}dt,   \eqno(3.14)  %(1spL)
$$

\ni where $\delta$ is the spike parameter. Comparing (3.14) with the expression under the integral of (3.1) one immediately sees that the special power $-\beta/2$ corresponding to the second choice of degenerate Virasoro primary like in (3.12) is present here. Therefore this integral and all its large $N$ limits satisfy PDEs of type (3.12), i.e.~with $\hbar=2/\beta$ as was found in the just cited references by probabilistic methods. Moreover, we see now that before taking any limit, the finite Laguerre (or Gaussian) $\beta$-ensembles in external field should satisfy a PDE of this type if we manage to express the corresponding left-hand side of (3.10) in terms of derivatives w.r.t.~the coupling parameters. %and in statistical context $a$ is the integer difference of dimensions of Gaussian rectangular $n\times m$ sample matrix $X$ whose covariance matrix $XX^{\dagger}$ is then $n\times n$ Wishart matrix. However, in general, the only restriction on $a$ is $a>-1$ needed for convergence of the integrals over the above density. %The hard edge limit arises here when $n\to \infty$, $n\delta \to x$, and the scaled spike parameter $x$ is fixed.
\par To summarize, introducing into a $\beta$-ensemble integral the factors corresponding to insertion of degenerate primary fields in CFT amounts to choosing a convenient generating function to probe the dependence of the integral on coupling parameters like $w_k$ in (3.1) or $t_l$ in (3.13), so that the modified integral (generating function) satisfies a relatively simple linear PDE of BPZ/FP/non-stationary Schr\"odinger type.
\par A key to revealing various symmetries and relations among different $\beta$-ensembles could become the remarkable duality formulas of Desrosiers~\cite{Der09},

$$
e^{-p_2(f)}\left\langle \prod_{j=1}^n\prod_{k=1}^N\left(s_j \pm i\sqrt\frac{2}{\beta}x_k\right) {}_0\F_0^{(2/\beta)}(x, 2f) \right\rangle_{x\in G\beta E_N} =
$$

$$
= e^{-p_2(s)}\left\langle \prod_{j=1}^n\prod_{k=1}^N\left(y_j \pm i\sqrt\frac{2}{\beta}f_k\right) {}_0\F_0^{(\beta/2)}(y, 2s) \right\rangle_{y\in G4/\beta E_n},   \eqno(3.15)  %(DG1) 
$$

$$
e^{-p_2(f)}\left\langle \prod_{j=1}^n\prod_{k=1}^N\left(s_j \pm x_k\right)^{-\beta/2} {}_0\F_0^{(2/\beta)}(x, 2f) \right\rangle_{x\in G\beta E_N} =
$$

$$
= e^{-p_2(s)}\left\langle \prod_{j=1}^n\prod_{k=1}^N\left(y_j \pm f_k\right)^{-\beta/2} {}_0\F_0^{(2/\beta)}(y, 2s) \right\rangle_{y\in G\beta E_n},   \eqno(3.16)  %(DG2) 
$$

\ni where ${}_0\F_0^{(\gamma)}(X,Y)$ is the hypergeometric function of two matrix arguments $X$ and $Y$ (see e.g.~\cite{Mac} or~\cite{F2010}), $p_2(s)=\sum_js_j^2$ etc., relating averaged products of $n$ characteristic polynomials in $s_j$-variables over Gaussian $\beta$-ensemble of size $N$ with external matrix source (external field) $f$ with eigenvalues $f_k$ to the ones of degree $N$ over dual $4/\beta$-ensemble of size $n$ with vectors $s$ and $f$ interchanged. There are similar formulas for Laguerre $\beta$-ensembles~\cite{Der09} involving function ${}_0\F_1$. Applying the transformation (2.10) one sees that in fact these formulas connect also various $\beta$-ensembles with non-Gaussian polynomial potentials in external fields. It would be interesting to extend these formulas e.g.~to the largest eigenvalue distributions of the corresponding RME, e.g.

$$
e^{-p_2(f)}\int_{-\infty}^t\dots\int_{-\infty}^t \prod_{j=1}^n\prod_{k=1}^N\left(s_j \pm i\sqrt\frac{2}{\beta}x_k\right) {}_0\F_0^{(2/\beta)}(x, 2f) \Delta^{\beta}(x)\prod_{i=1}^Ne^{-x_i^2}dx_i.   \eqno(3.17)  %(tDG)
$$

\ni Shifting the integration variables $x_i\to x_i+t$ one obtains an integral of the form

$$
e^{-p_2(f+t)}\int_{-\infty}^0\dots\int_{-\infty}^0 \prod_{j=1}^n\prod_{k=1}^N\left((s_j\mp it\sqrt\frac{2}{\beta}) \pm i\sqrt\frac{2}{\beta}x_k\right) {}_0\F_0^{(2/\beta)}(x, 2(f+t)) \Delta^{\beta}(x)\prod_{i=1}^Ne^{-x_i^2}dx_i.   \eqno(3.18)  %(tDGs)
$$

\ni Then, redefining the characteristic polynomial and external source variables by the opposite shift, $s_j\to s_j-t, f_k\to f_k-t$, one returns to the integral of the original form but over the intervals $(-\infty, 0]$ instead of $(-\infty, t]$. Thus, the problem reduces to the possibility of extension of the dualities between the integrals over the whole $\mathbb R^N$ to the ones over the single $N$-dimensional orthant. The idea of such shift of integration variables (in the opposite direction) was recently used in~\cite{NiSu14} to obtain information on gauge theory instanton partition function and a phase transition for it from the Gaussian unitary ensemble (GUE) largest eigenvalue probability and the Tracy-Widom distribution~\cite{TW-Airy} in its large matrix size limit.
%\par Let us compare the recent results of~\cite{DerLiu13, DerLiu11} based on~\cite{Der09} with the work~\cite{Nag11}.
\par Integrals of the form (2.1) also recently appeared~\cite{Gai, GaiTesch, AgEtAl11} in the string theory and AGT correspondence~\cite{AGT} between 2-dimensional Liouville CFT and 4-dimensional supersymmetric gauge theories, can be obtained from those with potential (3.9) by the limiting confluence procedure, when the Penner singularities $w_l$ merge. In CFT this corresponds to the introduction of degenerate or confluent primary fields, see e.g.~\cite{JimNagSu08}. This is quite similar to the confluence process of Garnier systems~\cite{GaussToP}, where the regular singular points merge leading from the Garnier systems to the confluent Garnier systems with irregular singularities~\cite{Kim89, Kaw03, Shim01}. 
 
\section{Results and conjectures for general $\beta$}

\subsection{Stationary Schr\"odinger equations and ODE/IM correspondence}

As a representative example, let us consider the system of~\cite{BLZ01, BLZ03}. In~\cite{BLZ01}, the authors proved the extended version of the earlier conjecture~\cite{DorTat99} about the relation between vacuum eigenvalues of CFT $Q$-operators (analogs of Baxter $Q$-operators for integrable lattice models and quantum spin chains, see e.g.~\cite{ODEIM} and references therein) introduced in~\cite{BLZ}, part II, and spectral determinants of one-dimensional Schr\"odinger equations (SE) on the half-line $x>0$ with homogeneous potential,

$$
\left(-\prt_{xx} + x^{2\alpha} + \frac{l(l+1)}{x^2}\right)\Psi(x) = E\Psi(x).   \eqno(4.1)  %(1)
$$

\ni The vacuum eigenvalues of $Q$-operators $Q_{\pm}(\lambda)$ are their eigenvalues at the highest weight state $|p\rangle$ of a representation of Virasoro algebra parameterized by a momentum-like parameter $p$, the highest weight $\Delta$ is given by $\Delta = p^2/\kappa + (c-1)/24$, and $c=1 - 6(1-\kappa)^2/\kappa$ is the central charge of the CFT. I.e.

$$
\langle p| Q_{\pm}(\lambda) |p\rangle = \lambda^{\pm 2\pi i p/\kappa}A_{\pm}(\lambda, p),   \eqno(4.2)  %(2)
$$

\ni where $\lambda$ is a spectral parameter. The analytical properties of $A_{\pm}(\lambda, p)$ as functions of $\lambda$ and $p$ were studied in~\cite{BLZ}. They turned out~\cite{DorTat99, BLZ01, DTCFT} to correspond to the properties of the spectral determinants

$$
D_{\pm}(E, l) = \prod_{n=1}^\infty\left(1 - \frac{E}{E^\pm_n}\right)   \eqno(4.3)  %(3)
$$

\ni of (4.1), where the ``plus" and ``minus" signs correspond to the even and odd eigenfunctions with the ordered eigenvalues $E^\pm_n$, respectively, with the degree and the ``angular momentum" parameter equal to

$$
\alpha = \frac{1}{\kappa} - 1,   \qquad  l = \frac{2p}{\kappa} - \frac{1}{2}.  \eqno(4.4)  %(4)
$$

\ni The relation shown to hold in~\cite{BLZ01} reads:

$$
A_{\pm}(\lambda, p) = D_{\pm}(\rho\lambda^2, 2p/\kappa - 1/2),   \qquad \rho = \left(\frac{2}{\kappa}\right)^{2-2\kappa}\Gamma^2(1-\kappa).   \eqno(4.5)  %(5)
$$

%\ni where

%$$
%\rho = \left(\frac{2}{\kappa}\right)^{2-2\kappa}\Gamma^2(1-\kappa).   \eqno(6)
%$$

\ni It holds for $\alpha>1$, i.e.~for $\kappa<1/2$ %. These relations, as we will see below, correspond to the range $\beta = 2\kappa \le 2$ 
(the range $1\le\beta\le 2$, i.e.~$\beta=1, 2$, should be reached by analytic continuation from the domain $\beta<1$, and the spectral determinants $D_{\pm}$ should then be defined differently from just (4.3), using the additional Weierstrass exponential factors in the infinite products, see e.g.~\cite{AbF2003} or~\cite{Vor87}). The proof of~\cite{BLZ01} proceeds as follows.
\par Assuming $\Re l > -3/2$, a solution $\psi^+(x,E)=\psi(x, E, l)$ of (4.1) is uniquely specified by the asymptotic at $x\to0$,

$$
\psi(x,E,l) \to \sqrt{\frac{2\pi}{1+\alpha}}\frac{x^{l+1}}{(2+2\alpha)^{(2l+1)/(2+2\alpha)}}\Gamma(1+(2l+1)/(2+2\alpha)) + O(x^{l+3}).  \eqno(4.6)  %(7)
$$

\ni It can be analytically continued outside $\Re l > -3/2$, and the function $\psi^-(x,E)=\psi(x, E, -l-1)$ is a linearly independent solution of the same equation for generic $l$ since their Wronskian $W[\psi^+, \psi^-]$ is equal to

$$
W[\psi^+, \psi^-] = \psi^+\prt_x\psi^- - \psi^-\prt_x\psi^+ = 2i(q^{l+1/2} - q^{-l-1/2}), \qquad q=e^{i\pi\kappa}=e^{\frac{i\pi}{1+\alpha}}.   \eqno(4.7)  %(8)
$$

\ni Then the asymptotics of (4.1) as $x\to+\infty$ are considered. There is a unique solution $\chi(x,E,l)$ which decays at large $x$, e.g.~normalized so that

$$
\chi(x,E,l) \to \frac{1}{x^{\alpha/2}}\exp\left(-\frac{x^{1+\alpha}}{1+\alpha} + O(x^{1-\alpha}\right).   \eqno(4.8) %(9)
$$

\ni Now the crucial role is played by the discrete symmetries of (4.1) given by the two transformations:

$$
\hat \Lambda: \quad x\to x,\ E\to E,\ l\to -l-1;  \qquad \hat \Omega: x\to qx,\ E\to q^{-2}E,\ l\to l.   \eqno(4.9)  %(10)
$$

\ni The transformation $\hat \Omega$ applied to $\chi(x,E,l)$ yields another linearly independent solution. With the choice of~\cite{BLZ01}, $\chi^-(x,E,l) = iq^{-1/2}\chi(qx, q^{-2}E, l)$, their Wronskian

$$
W[\chi, \chi^-] = 2.  \eqno(4.10)  %(11)
$$

\ni Then one matches the two asymptotics expanding the solutions $\psi^+$ and $\psi^-$ into the basis $\chi$, $\chi^-$, e.g.

$$
\psi^+ = C(E,l)\chi + D(E,l)\chi^-,   \eqno(4.11)   %(12)
$$

\ni where the notation $D(E,l)$ anticipates the result to be seen shortly. It is easy to see, that the action of the transformations (4.9) on the four solutions introduced is the following:

$$
\hat \Lambda\psi^{\pm} = \psi^{\mp}; \qquad \hat \Lambda\chi^{\pm} = \chi^{\pm},   \eqno(4.12)  %(13)
$$

$$
\hat\Omega\psi^{\pm} = q^{1/2 \pm (l+1/2)}\psi^{\pm}; \qquad \hat \Omega\chi^+ = -iq^{1/2}\chi^-,  \qquad  \hat \Omega\chi^- = -iq^{1/2}\chi^+ + u\chi^-,    \eqno(4.13)  %(14)
$$

\ni with some coefficient $u=u(E,l)$. From (4.13) and (4.11) it follows that

$$
C(E,l) = -iq^{-l-1/2}D(q^{-2}E, l),   \eqno(4.14)  %(15)
$$

\ni and applying (4.12) to (4.11) gives

$$
\psi^- = D(E, -l-1)\chi^- - iq^{l+1/2}D(q^{-2}E, -l-1)\chi^+.   \eqno(4.15)  %(16)
$$

\ni At last, taking the Wronskian $W[\chi^+, \psi^+]$ and using (4.11) and (4.10) one finds

$$
D(E,l) = \frac{1}{2}W[\chi^+, \psi^+].   \eqno(4.16)  %(17)
$$

\ni By (4.16), if $D(E, l)=0$, it means that $\psi^+\sim\chi^+$, which is an eigenfunction of discrete spectrum of (4.1), i.e.~$E$ must belong to the zeros of $D^+(E, l)$ (4.3), and vice versa. Both are entire functions of $E$. %($D(E,l)$ ?). 
Therefore also $\log(D^+(E,l)/D(E,l))$ is entire. By semiclassical asymptotics as $E\to\infty$ %(?) 
one finds that $\log(D^+(E,l)/D(E,l)) \to 0$ in this limit and hence $D^+(E,l) = D(E,l)$. From all the above facts one can establish (4.5) by showing that $D(E,l)$ satisfy the properties uniquely defining functions $A_+$~\cite{BLZ}. The bilinear relation

$$
q^{l+1/2}D(q^2E, l)D(E, -l-1) - q^{-l-1/2}D(E, l)D(q^2E, -l-1) = q^{l+1/2} - q^{-l-1/2}  \eqno(4.17)  %(18)
$$

\ni follows from combining (4.11), (4.14), (4.15), (4.7) and (4.10). It exactly corresponds the so-called quantum Wronskian relation which $A_\pm(\lambda, p)$ satisfy~\cite{BLZ}. Their matching analiticity conditions can be derived from (4.16) and WKB analysis of (4.1)~\cite{BLZ01, DTCFT}, we skip the details.
\par The correspondence between one-dimensional SE and quantum integrable models was extended to the excited states of the Baxter $Q$-operators in~\cite{BLZ03}. Then the SE $(-\prt_{xx} + V(x))\Psi(x) = E\Psi(x)$ arises with the following modified potential:

$$
V(x) = x^{2\alpha} + \frac{l(l+1)}{x^2} - 2\frac{d^2}{dx^2}\sum_{k=1}^L\ln(x^{2\alpha+2} - z_k),   \eqno(4.18) %(19)
$$

\ni where $z_k$ are $L$ pairwise different complex numbers satisfying $L$ algebraic Bethe ansatz type equations

$$
\sum_{j\neq k}^L\frac{z_k(z_k^2 + (3+\alpha)(1+2\alpha)z_kz_j + \alpha(1+2\alpha)z_j^2)}{(z_k-z_j)^3} - \frac{\alpha z_k}{4(1+\alpha)} + \frac{(2l+1)^2 - 4\alpha^2}{16(\alpha+1)} = 0.  \eqno(4.19)  %(20)
$$

\ni The SE with potential (4.18) is in fact equivalent to another one involving the ``circular polygon potentials" (see e.g.~\cite{AbF2003} about these potentials) under the natural change of variable already considered in~\cite{LitILW},

$$
z=x^{2\alpha+2} = x^{2/\kappa} \quad \Longrightarrow \quad x = z^{\kappa/2}.
$$

\ni In terms of variable $z$ the SE of~\cite{BLZ03} becomes

%$$
%\prt_x = \frac{2}{\kappa z^{\kappa/2 -1}}\prt_z,  \qquad  \prt_{xx} = \frac{4}{\kappa^2 z^{\kappa-2}}\prt_{zz} - \frac{2(\kappa-2)}{\kappa^2 z^{\kappa -1}}\prt_z,
%$$

$$
\left(-\prt_{zz} + \frac{\kappa-2}{2z}\prt_z + \frac{\kappa^2l(l+1)}{4z^2} + \frac{\kappa^2}{4z} + \sum_{k=1}^L\left(\frac{2}{(z-z_k)^2} + \frac{\kappa-2}{z(z-z_k)}\right) - \frac{\kappa^2E}{4}z^{\kappa-2}\right)\Psi(x(z)) = 0,   \eqno(4.20)  %(zBLZ)
$$

\ni and changing the wavefunction by the factor

$$
\Psi(x(z)) = z^{(\kappa-2)/4}\psi(z)
$$

\ni and rearranging terms in (4.20) yields the SE for the new function $\psi(z)$:

$$
\left(-\prt_{zz} + \frac{\kappa^2l(l+1) + (\kappa-2)(6-\kappa)/4}{4z^2} + \left(\frac{\kappa^2}{4} - \sum_{k=1}^L\frac{\kappa-2}{z_k}\right)\frac{1}{z} + \right.
$$

$$
\left. + \sum_{k=1}^L\left(\frac{2}{(z-z_k)^2} + \frac{\kappa-2}{z_k(z-z_k)}\right) - \frac{\kappa^2E}{4}z^{\kappa-2}\right)\psi(z) = 0.   \eqno(4.21)  %(LitBLZ)
$$

\ni This equation was obtained in~\cite{LitILW} with different notations, e.g.~$\kappa=-b^2$ where $b$ is the Liouville theory parameter considered in~\cite{LitILW} (the real parameter $b$ corresponds to $\beta<0$). In (4.21) we see indeed a polygon potential added to the original potential characterizing the ground state. In contrast to the case of nonstationary equation QPII considered below, here the appearance of the polygon potential is unrelated to the value of $\kappa$ which is tied with the degree $2\alpha$ of the ground state potential. 

\subsection{Simplest nonstationary cases -- quantum Painlev\'e equations}

These are Fokker-Planck (FP) or nonstationary Schr\"odinger equations (NSSE) in one "time" and one "space" variable, with Hamiltonians being the canonically quantized classical Painlev\'e Hamiltonians, i.e.~$H(x,p) \to H(x, \hbar\prt_x)$\footnote{the ususal ordering ambiguity of quantization is cured here by shifts of the free parameters of Painlev\'e equations}. Aside from CFT applications, their connection with $\beta$-ensembles first appeared for general beta from two very different sources. First, special cases of large matrix size $N$ limits of certain $\beta$-ensembles -- so-called soft edge and hard edge limits -- were found to satisfy the quantum PII~\cite{EdSut, RRV, BV1} and quantum PIII~\cite{RR08, RRZ} equations, respectively. In the above papers, however, the linear PDEs obtained were not identified with quantized Painlev\'e equations. This was done in the second source -- paper~\cite{Nag11}, which started from the canonical quantization of all Painlev\'e Hamiltonians and then showed that eigenvalue integrals of beta ensembles with special potentials were particular solutions of the quantum Painlev\'e (QP) equations. %It is very interesting to note that the values of $\beta$ in beta ensembles of~\cite{Nag11} correspond through the QP equations to the {\it dual} values $4/\beta$ in the ensembles where QP equations appeared from large $N$ limits. There is no clear explanation of this fact yet, but it must have something to do with the remarkable duality formulas for finite Gaussian and Laguerre beta ensembles revealed in~\cite{Der09}.
\par In~\cite{betaFP1}, we studied the quantum Painlev\'e II equation (QPII) from the point of view of possible exact quantum-classical correspondence -- we posed and partly solved the problem of finding a classical $2\times2$ matrix Lax pair such that one of its eigenvector components satisfies the QPII for general $\beta$ or $\kappa\equiv\beta/2$:

$$
\left(\kappa\prt_t + \prt_{xx} + (t-x^2)\prt_x\right)\F(t, x) = 0,   \eqno(4.22)  %(3.0)
$$

%$$
%(\kappa\prt_t + \sigma(t,x)\prt_{xx} + v(t,x)\prt_x + \alpha(t,x))\F_{\kappa}(t,x) = 0,   \eqno(2.1)
%$$

\ni driven in part by the fact of existence of the pair for two special values $\beta=2, 4$ (or $\kappa=1, 2$) found in~\cite{BV1} for the QPII and in~\cite{HBP3} for hard edge related QPIII,

$$
(\kappa t\prt_t + x^2\prt_{xx} + (ax - x^2 - 1/t)\prt_x)\F^H(t,x) = 0.   %\eqno(2.2) %(1)
$$

\ni The boundary conditions ensure that the solution $\F(t,x)$ is a probability distribution function. Moreover, its large $x$ asymptotic $F_\beta(t)$,

%$$
%\F^{(\beta)}(t,x) \to 0  \ \ \ \text{as } x \to -\infty, t < \infty,  \qquad   \F^{(\beta)}(t,x) \to 1 \quad \text{as } t, x \to \infty \text{ together},
%$$

$$
\F(t,x) \to F_\beta(t)  \quad \text{as } x \to +\infty,    \eqno(4.23)  %(1.2) % t \text{ finite}.  \eqno(1.2)
$$

\ni is the Tracy-Widom distribution ($TW_\beta$) for general $\beta$. The celebrated distributions $TW_2$~\cite{TW-Airy}, $TW_1$ and $TW_4$~\cite{TW-OrtSym} of RMT with unitary, orthogonal and symplectic symmetry, respectively, giving it the name universally appear in certain asymptotics of random particle system processes, see e.g.~\cite{Fer13} and references therein. 
\par We set up to find a Lax pair of the form

$$
\prt_x\left(\begin{array}{c}\F \\ G \end{array}\right) = L\left(\begin{array}{c}\F \\ G \end{array}\right),  \ \ \ \ \ \   \prt_t\left(\begin{array}{c}\F \\ G \end{array}\right) = B\left(\begin{array}{c}\F \\ G \end{array}\right),  \eqno(4.24)  %(2.2)   %\eqno(2.5)
$$

\ni where we denoted

$$
L = \left(\begin{array}{cc} L_1 & L_+ \\ L_- & L_2 \end{array}\right),   \ \ \ \ \ B = \left(\begin{array}{cc} B_1 & B_+ \\ B_- & B_2 \end{array}\right).
$$

\ni Then, eliminating $G$ from the first components of these equations, one obtains another, first-order, PDE for $\F$:

$$
\prt_t\F - b_+\prt_x\F + b_1\F = 0,   \eqno(4.25)  %(2.4)
$$

\ni where we denoted

$$
b_+ = \frac{B_+}{L_+}, \ \ \ \ \ b_1 = b_+L_1-B_1.   \eqno(4.26)  %(2.5)
$$

\ni Eliminating $\prt_t\F$ from (4.22) and (4.25), one sees that $\F$ satisfies also an ODE in $x$:

$$
(\prt_{xx} + (t-x^2+\kappa b_+)\prt_x - \kappa b_1))\F = 0,  \eqno(4.27)  %(2.6)
$$

\ni which amounts to the effective separation of variables in QPII. All this is in fact consistent in a greater generality~\cite{betaFP1}, i.e.~for an FP equation with general differentiable coefficients in place of QPII. It depends only upon the solvability of a closed governing system of two nonlinear PDEs for the functions $P=-\kappa b_+$ and $b=\kappa b_1$, which in QPII case reads\footnote{This system with $v=0$ to the best of our knowledge first appeared in~\cite{BC69} where it was used to find similarity solutions to the heat equation.}($v(t,x)=t-x^2$):

$$
\kappa\prt_t\left(P - v\right) + \prt_x\left(\prt_xP + P(P - v) + 2b\right) = 0,   \eqno(4.28)  %(3.1)  %(28)
$$

$$
\kappa\prt_tb + \prt_{xx}b + v\prt_xb = -2b\prt_xP.   \eqno(4.29)  %(3.2)  %(29)
$$

\ni We found~\cite{betaFP1} an explicit solution of this system for all integer $\kappa$ (i.e.~even $\beta$), using an intuition from the known simplest cases $\kappa=1, 2$~\cite{BV1, HBP3} and from a consideration of quite similar problem for all Painlev\'e equations, but not considering general $\beta$, in~\cite{ZZFP}. In~\cite{ZZFP}, the importance of poles in zero-curvature equations was stressed and solutions with one and two poles, corresponding to Painlev\'e equations were found, starting from an FP equation like (4.22). Their solutions in fact correspond to $\kappa=1, 2$, respectively. Our solution for all $\kappa\in\mathbb N$ has $\kappa$ poles:

$$
b_+(t,x) = -\frac{1}{\kappa}\sum_{k=1}^{\kappa}\frac{1}{x-Q_k(t)},   \eqno(4.30) %(3.7)  %(34)
$$

$$
 2b_1(t, x) = \frac{1}{\kappa}\sum_{k=1}^\kappa\frac{\kappa Q_k' + t - Q_k^2 - 2R_k}{x-Q_k} - \frac{1}{\kappa}\sum_{k=1}^\kappa Q_k - \frac{1}{\kappa}\left(\frac{t^2}{2} + U(t)\right),   \eqno(4.31)  %(3.14)  %(41)
$$

\ni where we denoted

$$
R_k = \sum_{j\neq k}^N\frac{1}{Q_k - Q_j},   \eqno(4.32)  %(3.15)  %(42)
$$

\ni function $U(t)$ is defined by

$$
\kappa U'(t) = -\sum_{k=1}^{\kappa}Q_k^2,   \eqno(4.33)  %(3.27)    %(3.37)  %(64)
$$

\ni and the poles $Q_k(t)$ satisfy equations of motion with Calogero interaction in external time-dependent cubic (``Painlev\'e II implying") potential,

$$
\kappa^2 Q_k'' = -2Q_k(t-Q_k^2) + \kappa-2 - \sum_{j\neq k}\frac{8}{(Q_k-Q_j)^3},  \eqno(4.34)  %(3.29)   %(3.36)  %(63)
$$

\ni which, considered together with (4.33), have $\kappa$ first integrals

$$
\frac{(\kappa Q_k')^2}{2} + tQ_k^2 - \frac{Q_k^4}{2} - (\kappa-2)Q_k - \sum_{j\neq k}^{\kappa}\frac{2}{(Q_k - Q_j)^2} + U(t) - 
$$

$$
- \sum_{j\neq k}^{\kappa}\frac{\kappa Q_k' + \kappa Q_j'}{Q_k - Q_j} + \sum_{j\neq k}^{\kappa}\sum_{l\neq k,j}^{\kappa}\frac{2}{(Q_k - Q_j)(Q_j - Q_l)} = 0,  \eqno(4.35)  %(3.30)   %(3.x6)
$$

\ni (which can be written more concisely if one uses the exchange operators between $Q_k$ particles~\cite{betaFP1}). The functions $Q_k(t)$ are in many respects similar to the coordinates of Garnier system~\cite{GaussToP}, e.g.~in their origin as apparent singularities of equation (4.27). By comparing with the considerations of the previous subsection, one observes that the equations (4.35) play here the role of {\it non-stationary Bethe ansatz equations}.
\par Thus, the exact quantum-classical correspondence is established for quantum Painlev\'e II -- explicitly for $\kappa\in\mathbb N$ and conjecturally, but plausibly, since the system (4.28), (4.29) has a Laurent series solution for all $\kappa$. The found integrable structure has already allowed us to obtain a Painlev\'e II representation for $TW_6$ ($\kappa=3$)~\cite{betaTWk3} which was beyond the classical integrable theory before. {\bf In fact, see sections 3, 4.3, 6, in~\cite{betaTWk3} we implicitly used the irregular ``5-point conformal block with one degenerate primary" $\sim\F(t,x)$ to gain information on the irregular ``4-point conformal block" $\sim\F_0(t)$}. Here we want to present some further facts and conjectures about the system for various $\kappa$.
\ni Recall the ODE in $x$ (4.27) that a solution of QPII must satisfy if the governing system has a solution. For $\kappa\in\mathbb N$ this is certainly the case. The ODE can be brought to a form of stationary Schr\"odinger equation by making the first derivative drop out:

$$
\F = \Psi Y^{1/2}e^{-1/2(tx-x^3/3)} = \Psi e^{1/2\int^xP(t,x)dx}e^{-1/2(tx-x^3/3)},
$$

\ni so that function $\Psi(x;t)$ satisfies

$$
\prt_{xx}\Psi - V\Psi = 0,    \qquad  V = b + \frac{(P-t+x^2)^2}{4} - \frac{\prt_x(P-t+x^2)}{2},    \eqno(4.36)  %(xODE) 
$$

%$$
%\kappa \prt_t\Psi + P\prt_x\Psi - \frac{\prt_x P}{2}\Psi = 0.    \eqno(1-ord)  
%$$

\ni which is explicit for $\kappa\in\mathbb N$:

$$
V(x) = \frac{3}{4}\sum_{k=1}^{\kappa}\frac{1}{(x-Q_k)^2} + \frac{1}{2}\sum_{k=1}^{\kappa}\frac{\kappa Q_k' - R_k}{x-Q_k} - \frac{U(t)}{2} + \frac{(\kappa-2)x}{2} - \frac{tx^2}{2} + \frac{x^4}{4}.  \eqno(4.37)  %(xODEin)
$$

\ni This potential is a sum of quartic potential for non-symmetric anharmonic oscillator (so has an irregular singularity at $x=\infty$) and the potential arising in studies of conformal mappings of circular polygons~\cite{AbF2003}. The appearance of the circular polygon potentials seems to be generic for conformal theory related problems, recall e.g.~the higher order eigenvalues of quantum transfer operators~\cite{BLZ03} from the previous subsection. This genericity finds a natural explanation in the theory of isomonodromic deformations where the vertices of the polygons appear as {\it apparent singularities} i.e.~the singular points of the coefficients of an ODE in the $x$-complex plane such that its solutions are meromorphic at these points~\cite{GaussToP, FIKN, DubMa07}. Indeed, all the $Q_k$ in (4.37) are apparent singularities. Moreover, for general $\kappa$, all the possible singularities of the function $P(t,x)$ are simple poles and they lead to such apparent singularities for the equation (4.36).
\par For general $\kappa$, the governing system (4.28), (4.29) can be rewritten in terms of $P$ and the Schr\"odinger potential $V$ (4.36) as follows:

$$
\kappa\prt_tV + P\prt_xV + 2V\prt_xP - \frac{\prt_{xxx}P}{2} = 0,   \eqno(4.38)  %(PBconn)
$$

$$
\kappa\prt_tP + 2\prt_{xx}P + P\prt_xP + 2x(t-x^2) - (\kappa-2) + 2\prt_xV = 0.   \eqno(4.39)  %(C-law)
$$

\ni Comparing with CSM-related governing systems arising in CFT, see section 5.2, one can see that the main difference is in the additional time-dependent free term $2x(t-x^2) - (\kappa-2)$ here reflecting the fact that we deal with Calogero-like system in time-dependent external field.
\par Also, for general $\kappa$, the logarithmic $x$-anti-derivative of $P$, let us denote it $Y$, i.e.

$$
P = \frac{\prt_xY}{Y},
$$

\ni satisfies a bilinear PDE of Hirota-like form,

$$
Y(\kappa\prt_t + \prt_{xx})^2Y - ((\kappa\prt_t + \prt_{xx})Y)^2 - 2\prt_xY\prt_x(\kappa\prt_t + \prt_{xx})Y + 2\prt_{xx}Y(\kappa\prt_t + \prt_{xx})Y + 
$$

$$
+ Y(\kappa\prt_tf(v)\cdot Y + \prt_xf(v)\cdot \prt_xY) + 2f(v)(Y\prt_{xx}Y - (\prt_xY)^2) = 0,    \eqno(4.40)  %(HirY)
$$

\ni where

$$
f(v) = -\int_0^x\prt_tv(t,z)dz - \prt_xv - \frac{v^2}{2} = -(\kappa-2)x - \frac{(t-x^2)^2}{2}.
$$

\ni For $\kappa\in\mathbb N$, it has polynomial solutions, $Y= \prod_{k=1}^\kappa(x-Q_k(t))$, in terms of the above functions $Q_k(t)$. The explicit Lax pair found for $\kappa\in\mathbb N$, reads\footnote{The Lax pair is not unique, in general it has two arbitrary functions of $x$ and $t$, but the presented form seems convenient for general $\kappa\in\mathbb N$, being polynomial in $x$.}~\cite{betaFP1}

%$$
%L = \left(\begin{array}{cc} L_1 & L_+ \\ L_- & L_2 \end{array}\right) = \left(\begin{array}{cc} \frac{1}{2}(-v+L_d) & L_+ \\ -\frac{1}{2L_+}(\kappa B_d + \prt_xL_d + L_d^2/2 + f(v)) & \frac{1}{2}(-v-L_d) \end{array}\right),   %\eqno(1.5)  
%$$

$$
L = \left(\begin{array}{cc} \frac{1}{2}(-v+L_d) & Y \\ -\frac{1}{2Y}(\kappa B_d + \prt_xL_d + L_d^2/2 + f_v) & \frac{1}{2}(-v-L_d) \end{array}\right),   %\eqno(1.5)  
$$

%$$
%B = \left(\begin{array}{cc} B_1 & B_+ \\ B_- & B_2 \end{array}\right) = \left(\begin{array}{cc} \frac{1}{2}\left(-x+\frac{U(t)+t^2/2}{\kappa}-\frac{\phi'}{\phi} + B_d\right) & -\frac{\prt_xL_+}{\kappa} \\ -\frac{2L_-\prt_xL_+ + \kappa \prt_tL_d - \kappa \prt_xB_d}{2\kappa L_+} & \frac{1}{2}\left(-x+\frac{U(t)+t^2/2}{\kappa}-\frac{\phi'}{\phi} - B_d\right)  \end{array}\right),   %\eqno(1.6)
%$$

$$
B = \left(\begin{array}{cc} \frac{1}{2}\left(-x+\frac{U(t)+t^2/2}{\kappa} + B_d\right) & -\frac{\prt_xY}{\kappa} \\ -\frac{2L_-\prt_xY + \kappa \prt_tL_d - \kappa \prt_xB_d}{2\kappa Y} & \frac{1}{2}\left(-x+\frac{U(t)+t^2/2}{\kappa} - B_d\right)  \end{array}\right),   %\eqno(1.6)
$$

\ni where $v= t-x^2$, 

%$$
%L_+ = \phi(t)\prod_{k=1}^{\kappa}(x-Q_k(t)),   %\eqno(1.8)    %(4.1)   
%$$

$$
L_d = -Y\cdot\sum_{k=1}^{\kappa}\frac{\kappa Q_k' - 2R_k}{(x-Q_k)\prod_{j\neq k}^{\kappa}(Q_k-Q_j)} = -\sum_{k=1}^{\kappa}(\kappa Q_k' - 2R_k)\prod_{j\neq k}^{\kappa}\frac{x-Q_j}{Q_k-Q_j},  % \eqno(1.9)   %(4.9)
$$

%$$
%\kappa B_d = \kappa \phi'(t)/\phi + \sum_{k=1}^{\kappa}\frac{\kappa Q_k' - 2R_k}{x-Q_k}\left(\sum_{l=1}^{\kappa}\frac{\prod_{j\neq l}^{\kappa}(x-Q_j)}{\prod_{j\neq k}^{\kappa}(Q_k-Q_j)} - 1\right),  % \eqno(1.10)    %(4.10)
%$$

$$
\kappa B_d = \sum_{k=1}^{\kappa}\frac{\kappa Q_k' - 2R_k}{x-Q_k}\left(\sum_{l=1}^{\kappa}\frac{\prod_{j\neq l}^{\kappa}(x-Q_j)}{\prod_{j\neq k}^{\kappa}(Q_k-Q_j)} - 1\right),  % \eqno(1.10)    %(4.10)
$$

$$
f_v = -\frac{x^4}{2} - tx^2 + (\kappa-2)x - U(t).  % + \frac{\kappa \phi'(t)}{\phi}\right).  %\eqno(1.11)   %(4.12)
$$

\par Another example with similarities and important differences from the foregoing, is the system~\cite{GiVes09} originating from NSSE%of Calogero particles in external (sextic) potential,

$$
i\prt_t\psi = \frac{1}{2}\left(-\prt_{xx} + x^6 - \nu x^2 + \frac{l(l+1)}{x^2}\right)\psi = 0.   \eqno(4.41)  %(frQP4)
$$

\ni Here, unlike above, the potential of the NSSE does not depend on time $t$, an important difference. Starting with this equation and using Darboux transformations by its quasirational solutions of the form $\psi^{(0)} = \sum_j^Mc_j\psi_j(x)\exp(-i\lambda_jt/2)$, where $\psi_j$ is quasirational solution of the corresponding SE with eigenvalue $\lambda_j$, for special values of $\nu$ and $l$, the authors derived a new NSSE with an additional potential having $m$ double poles, so that the total (time-dependent) potential is

$$
V(t, x) =  \frac{1}{2}\left(x^6 - \nu x^2 + \frac{l(l+1)}{x^2}\right) - 2\prt_{xx}\ln W_m(\psi^{(0)}_n)  % =  \frac{1}{2}\left(x^6 - \nu x^2 + \frac{l(l+1)}{x^2}\right) - 2\sum_{k=1}^{n_p or m?}\frac{\gamma_k(\gamma_k+1)}{(x-y_k)^2}
$$

\ni in terms of the Wronskian $W_m$ of $m$ independent quasirational solutions $\psi^{(0)}_n$, $n=1,\dots,m$, of (4.41) found in~\cite{GiVes09}. The function

$$
\psi^{(m)}(t,x) = x^{\mu}\frac{\prod_{j=1}^{n_z}(x-x_j(t))^{\alpha_j}}{\prod_{k=1}^{n_p}(x-y_k(t))^{\gamma_k}}e^{-x^4/4}\exp(if(t)),  \qquad  \mu = -l \text{ or } l+1,
$$

\ni where all $\alpha_j\in\mathbb  N$, $\gamma_k\in\mathbb  N$, similar in form to the solutions of the SE, solves the obtained NSSE for $\mu\in\mathbb Z$, $\nu=3+2\mu+2N$, %$\nu_{eff}=3+2\mu+2\sum_j\alpha_j-2\sum_k\gamma_k$, 
$N\in\mathbb Z$, if $\sum_j\alpha_j-\sum_k\gamma_k=N-3m-\mu$, the constraints $\sum_j\alpha_jx_j=\sum_k\gamma_ky_k$ and $\sum_j\alpha_j/x_j=\sum_k\gamma_k/y_k$ are satisfied, and, finally, the functions $x_j(t)$ and $y_k(t)$ satisfy the time-dependent Bethe ansatz type equations ensuring also the absence of monodromy for the solutions $\psi^{(m)}$, 

$$
-i\dot x_j + \sum_{k\neq j}^{n_z}\frac{\alpha_k}{x_j-x_k} - \sum_{k=1}^{n_p}\frac{\gamma_k}{x_j-y_k} - x_j^3 + \frac{\mu}{x_j} = 0,
$$

$$
i\dot y_k - \sum_{j=1}^{n_z}\frac{\alpha_j}{y_k-x_j} - \sum_{j\neq k}^{n_p}\frac{\gamma_j}{y_k-y_j} + \gamma_ky_k^3 - \frac{\mu}{y_j} = 0,
$$

$$
\sum_{j=1}^{n_z}\alpha_jx_j^2 - \sum_{k=1}^{n_p}\gamma_ky_k^2 + \dot f = 0
$$

\ni (compare with (4.35) and (4.33) !). In the case $\mu=0$, $\alpha_j=\gamma_k=1$ for all $j$ and $k$, the $x_j$ and $y_k$ decouple into similar independent subsystems of equations of motion,

$$
\ddot x_j(t) = -3x_j^5 - (2(n_z-n_p)-3)x_j + \sum_{l\neq j}^{n_z}\frac{2}{(x_j-x_l)^3},
$$

$$
\ddot y_k(t) = -3y_k^5 - (2(n_z-n_p)+3)y_k + \sum_{l\neq k}^{n_p}\frac{2}{(y_k-y_l)^3}.
$$

\ni A similarity with our equations (4.34) here is especially transparent as is the important difference due to the absence of terms explicitly depending on $t$. This system in fact can be considered as a peculiar ``remnant" of quantum Painlev\'e IV (QPIV). Indeed, starting with the form of QPIV written down e.g.~in~\cite{Nag11} and making simple change of $x$-variable $x\to x^2$ there and then changing the dependent function $\Psi$ by $\Psi=x^{1/2}\psi$ to remove the first $x$-derivative, one arrives at the equation of the form

$$
\left(\kappa\prt_t + \frac{1}{4}\prt_{xx} - \frac{x^6}{4} - t\frac{x^4}{2} + (n-\kappa-1/2-t^2/4)x^2 + \frac{\kappa(2-\kappa)-3/4}{4x^2} - (\alpha+\kappa/2)t\right)\psi = 0,   \eqno(4.42)  %(QP4)
$$

\ni with some parameters $n$ and $\alpha$ ($n$ is the integer number of a $\beta$-ensemble eigenvalues in~\cite{Nag11}), which contains the substantial additional term $\sim tx^4$ as compared to (4.41). This implies that the system of~\cite{GiVes09} might be considered from the more general QPIV point of view. The exact relation between them, however, needs a further investigation.
%The results of~\cite{GiVes09} imply that the simple solutions $\psi^{(m)}$ above obtained by Darboux construction give only a proper subset of possible monodromy-free solutions and potentials with sextic growth at infinity.
%\ni The important difference with the Quantum Painlev\'e II case~\cite{betaFP1} considered below is that the Hamiltonian () is independent of time while the time dependence of the functions () below is crucial. This does not allow to immediately extend the results of~\cite{GiVe09} to the QPII case but still some insights from~\cite{GiVe09} might be useful for finding the eigenfunctions of () related to QPII, this is a work in progress.

\subsection{Quantum Painlev\'e equations in CFT}

%Besides the mentioned early abstract considerations of~\cite{Sul94, Slav00, Sul08}, 
\par The quantum Painlev\'e (QP) equations appeared naturally in the studies of quantum integrable systems and CFT. The first detailed consideration of QPVI (quantum Painlev\'e VI) in this context seems to date back to 1993, where it appeared in~\cite{EtKir93} in representation-theoretic studies of quantum affine algebras and Wess-Zumino-Novikov-Witten (WZNW) quantum field theory. %\footnote{The authors apparently did not realize its relation with Painlev\'e VI then and considered only the parameterization in terms of elliptic functions}. 
More recently QPVI surfaced in the studies of Liouville CFT correlators in~\cite{FLNO} where both rational and elliptic parameterizations were considered and explicit transformation between them written down. Moreover, exact solutions of QPVI for special values of the parameters were found, related with the elliptic or algebraic solutions of Painlev\'e VI. Liouville CFT considered by~\cite{FLNO} has central charge $c>25$ (i.e.~$\beta<0$), but results for CFT with $c<1$ related to the $\beta$-ensembles can be obtained from it by careful analytic continuation. By the principles of CFT based on representation theory of Virasoro algebra (2.5), the general 4-point correlation function of primary operators has an expression

$$
\langle V_{\alpha_1}(z_1, \bar z_1)V_{\alpha_2}(z_2, \bar z_2)V_{\alpha_3}(z_3, \bar z_3)V_{\alpha_4}(z_4, \bar z_4) \rangle = 
$$

$$
= \prod_{i<j}|z_i-z_j|^{\gamma_{ij}}\int C(\alpha_1,\alpha_2, iP + Q/2)C(Q/2 - iP, \alpha_3, \alpha_4)\left| \F_P(\{\alpha_i\}; t) \right|^2\frac{dP}{2},    \eqno(4.43)  %(4corr)
$$

\ni where $Q=b+1/b$ with $b$ being the coupling parameter of the Liouville CFT ($b^2 = -\beta$), $\gamma_{ij}$ are certain combinations of conformal dimensions (weights), $C(\alpha, \gamma, \delta)$ are the structure constants of conformal algebra (which determine the CFT 3-point correlation functions), $\F_P(\{\alpha_i\}; t)$ called conformal blocks are in general known only as infinite series in the cross-ratio

$$
t = \frac{(z_1-z_2)(z_3-z_4)}{(z_1-z_3)(z_2-z_4)}
$$

\ni invariant w.r.t.~the global conformal i.e.~projective transformations, and integration is w.r.t.~the momentum parameter $P$ which determines the intermediate scaling dimensions $\Delta=P^2+Q^2/4$ in the given scattering cross-channel. If one of the four primaries is the so-called degenerate field, i.e.~$V_\alpha$ with a scaling dimension of the form

$$
\alpha = \alpha_{mn} = -\frac{mb}{2} - \frac{n}{2b}   \eqno(4.44)  %(ddim)
$$

\ni with integer $m$ and $n$, then the correlator (4.43) satisfies a PDE which is of second order e.g.~for $m=1, n=0$ (how it arises we saw with the $\beta$-ensemble model in section 3) and in the considered case is equivalent to a Gauss hypergeometric equation. The authors of~\cite{FLNO} considered the case $n=0$ and general $m$, which they studied with the help of a 5-point correlator with one degenerate field $V_{1/2b}$, 

$$
\langle V_{1/2b}(z)V_{\alpha_1}(0)V_{\alpha_2}(1)V_{\alpha_3}(\infty)V_{\alpha_4}(t) \rangle,
$$

\ni where the antiholomorphic dependence identical to the holomorphic one is suppressed in the notation. The last function satisfies a QPVI equation, which in the elliptic parameterization takes Schr\"odinger form

$$
\left(\frac{4i}{\pi b^2}\prt_{\tau} + \prt_{uu} - \sum_{k=1}^4s_k(s_k+1)\mathcal P(u-\omega_k) + \sum_{k=1}^3s_k(s_k+1)\mathcal P(\omega_k)\right)\Psi(u, \tau),   \eqno(4.45) %(eQP6)
$$

\ni where $\mathcal P(y)$ is the Weierstrass elliptic function,

$$
\tau = i\frac{K(1-t)}{K(t)}, \qquad u = \frac{\pi}{4K(t)}\int_0^{\frac{z-t}{t(z-1)}} \frac{dr}{\sqrt{r(1-r)(1-tr)}},     \eqno(4.46)  %(par)
$$

\ni $K(t)$ is the elliptic integral of the first kind, 

$$
\qquad K(t) = \frac{1}{2}\int_0^1 \frac{dr}{\sqrt{r(1-r)(1-tr)}},
$$

\ni $\omega_k$ are half-periods

$$
\omega_1 = \frac{\pi}{2}, \qquad \omega_2 = \frac{\pi\tau}{2}, \qquad \omega_3 = \omega_1+\omega_2, \qquad \omega_4=0,
$$

\ni $s_k$ are related to parameters $\alpha_k$ as

$$
\alpha_k = \frac{Q}{2} - \frac{b}{2}\left(s_k+\frac{1}{2}\right),    \eqno(4.47)  %(spar)
$$

\ni and the 5-point correlator is given in terms of $\Psi$ and several factors containing elliptic $\Theta_1$ function,

$$
\langle V_{1/2b}(z)V_{\alpha_1}(0)V_{\alpha_2}(1)V_{\alpha_3}(\infty)V_{\alpha_4}(t) \rangle = (z(z-1))^{1/b^2}\frac{(z(z-1)(z-t))^{1/4}}{(t(t-1))^{\frac{8\Delta(\alpha_4)+1}{12}}}\frac{(\Theta_1(u))^{1/b^2}}{(\Theta'(0))^{(1+1/b^2)/3}}\Psi(u, \tau).
$$

\ni The authors noted that for special values of parameters $s_k$,

$$
s_k = m_k + \frac{2n_k}{b^2}, \quad m_k, n_k \in \mathbb Z,
$$

\ni the general solution of the QPVI (4.45) can be obtained from the general solution of the corresponding heat equation, i.e.~(4.45) without the elliptic potential. Using this, they found explicitly exact solutions of QPVI for $s_1=s_2=s_3=0, s_4 = m + 2n/b^2$ in terms of $m+n$-dimensional integrals of elliptic functions and obtained some 4-point conformal blocks as limits from them. This QPVI for integer values of all $s_k$ also appeared in the studies of special symmetry point of eight-vertex model~\cite{BM05, BM0610} where also some special solutions have been obtained in terms of elliptic functions and their relation with elliptic solutions of Painlev\'e VI was revealed. For integer $s_k$ the potential is a finite gap potential. More general family of special solutions for the whole 4-dimensional lattice of integer $s_k$ and the corresponding Painlev\'e VI $\tau$-functions have later been found in~\cite{Ros13}. Thus, for QPVI with any value of $\kappa$ (or $b$), the special values of Painlev\'e VI related four parameters admit relatively simple exact solutions, just like for classical Painlev\'e VI itself.
\par Exactly as with classical Painlev\'e equations, it is possible to obtain all the other QP equations from QPVI by confluence of singularities procedure. This way e.g.~the exponential form of QPIII was obtained from (4.45) in~\cite{MMM11}. 

\section{Special cases $\beta=2 (c=1)$ and $c\to\infty$ ($\beta \to 0$ or $\beta\to\infty$)}

In these two cases many stronger results have been obtained than for general $\beta$ (or central charge $c$). The main point we wish to stress is that most of the results described below must be possible to extend to the general $\beta$ and related theories by means of the exact quantum-classical correspondence like the one we revealed in~\cite{betaFP1} and discuss in section 4.2. Indeed, Garnier systems (both regular and confluent) can be naturally considered as compatibility conditions of linear systems of PDEs. The linear PDEs involved are equivalent to the BPZ (or Knizhnik-Zamolodchikov (KZ)~\cite{KZ}) equations of a (confluent) CFT for special values of central charge $c$, i.e.~effectively $\beta$, usually for $\beta=2$, but also in the limits as $\beta\to\infty$ or $\beta\to0$. We propose that a similar equivalence may hold for general values of $\beta$, with some type of (modified) Garnier or more general isomonodromy systems. A concrete example of this we encountered studying quantum Painlev\'e II equation (QPII) related to the soft edge large size limit of $\beta$-ensembles, for all even integer $\beta$~\cite{betaFP1} and conjecturally (not completely explicitly yet) for the other $\beta$, see section 4.2. What leads us to this conjecture is the existence of classical integrable structure for general $\beta$ implied by the results of~\cite{betaFP1} as well as the appearance and crucial importance of a number of {\it apparent singular points} of differential equations, both in BPZ equations and in isomonodromy systems~\cite{GaussToP, FIKN, DubMa07}. The addition of a sufficient number of apparent singularities ensures the solvability of a general Riemann-Hilbert problem of finding a Fuchsian ODE with given singularities and their monodromies~\cite{GaussToP, FIKN}.

\subsection{Results for $\beta=2$ ($c=1$) -- traditional exact classical integrability}

The isomonodromic deformations of Fuchsian ODEs leading to the Garnier systems are excellently described in the book~\cite{GaussToP}. E.g.~Heun equation with one additional apparent singularity at the value of the {\it accessory parameter} $q$,

$$
\prt_{xx}y + \left(\frac{c}{x} + \frac{d}{x-1} + \frac{e}{x-t} - \frac{1}{x-q}\right)\prt_xy + \left(\frac{ab}{x(x-1)} + \frac{ht(t-1)}{x(x-1)(x-t)} + \frac{pq(q-1)}{x-q}\right),   \eqno(5.1)  %(P6xODE)
$$

\ni where $p(t)$ is the classical momentum for the coordinate $q(t)$ and $h(q,p,t)$ is the associated Painlev\'e VI Hamiltonian,

$$
h = \frac{ q(q-1)(q-t)p^2 + ((c-1)(q-1)(q-t) + (d-1)q(q-t) + eq(q-1))p + ab(q-t) }{t(t-1)},  %\eqno(5.2)  %(hP6)
$$

\ni describes the isomonodromic deformations of Heun equation itself leading to Painlev\'e VI satisfied by function $q(t)$. Equation (5.1) is the equation of the type (4.27) for QPVI with $\hbar = \kappa =1$ which itself is the nonstationary FP equation

$$
\hbar\prt_t y = h(x, \hbar\prt_x, t)y
$$

\ni with $h(q, p, t)$ above. All types of the confluent Heun equations with one apparent pole added similarly lead to the other Painlev\'e equations as written out e.g.~in~\cite{Slav00}. As is clear now from the results of~\cite{betaFP1}, this is, for QPVI in place of QPII, exactly the simplest case $\kappa=\beta/2=1$ (i.e.~$\hbar=1$) of the series of cases with $\kappa\in\mathbb N$ where the classical Lax pairs related to ODEs with $\kappa$ such apparent singularities are available. Some confluent Garnier systems, with two or more ``time" variables like $t$ above, were studied e.g.~in~\cite{Kim89, Kaw03, Shim01}. Generic determinantal solutions of Garnier system have been recently found in~\cite{Mano12}.
\par The idea of quantized Painlev\'e equations in the sense considered here, i.e.~as FP or NSSE equations in one space and one time dimensions, appeared in the papers of Suleimanov (e.g.~\cite{Sul9408}), long before the recent prominent applications. He used the scalar Garnier Lax pair for Painlev\'e II (PII) to work out the connection of QPII with classical integrability and PII itself. A restricted version of governing system~\cite{betaFP1} also appeared there. The connection of QPII with PII works this way only for $\hbar=1$ ($\beta=2$). %and rescaling the variables, e.g.~used in~\cite{Sul9408}, does not change effective $\hbar$. %just like rescaling of the coupling constant in classical CSM model. 
Later the relations between Painlev\'e equations and NSSE satisfied by solutions of the associated linear problem were developed in much detail in the works of Zabrodin and Zotov~\cite{ZZQPC, ZZFP}, using various parameterizations (trigonometric/exponential and elliptic besides usual rational) and ideas of Painlev\'e-Calogero correspondence~\cite{PCLevOlTak} bringing the classical Painlev\'e equations into the form of standard Newton equations of motion in special Calogero-like but time-dependent potentials. There again effectively $\hbar=\kappa=1$, except for the case of two poles in~\cite{ZZFP}, which may correspond to $\kappa=2$, see section 4.2, or $\kappa=1/2$ depending on the type of PDE like in section 3.
\par Other type of relevant results for $\beta=2$ came from random matrix theory and its applications in statistics studying large random samples of large number of (correlated) variables. E.g.~sample covariance matrix eigenvalue p.d.f.~leads to the integrals similar in form to the left-hand side of (3.17) with $n=0$. For $\beta=2$, phase transition from distribution $TW_2$ through critical 2-variable distribution $\F(t,x)_{\beta=2}$ of section 4.2 to Gaussian distribution for several largest eigenvalues was rigorously shown and quantified in~\cite{BBP, BaikP06}. Using integrable structure, e.g.~Darboux transformations,~\cite{BerCaf14} extended these considerations to express multiparameter critical distributions, related to certain determinantal random point processes and multimatrix Dyson BMs, in terms of simpler distributions like $\F(t,x)_{\beta=2}$. Some results, e.g.~the BBP\cite{BBP} phase transition and description of critical distributions for all $\beta$, have already been achieved for the integrals of the type (3.15)~\cite{DerLiu13}, which is another indication that further generalizations to all $\beta$ are possible.

%$$
%\Prob(\lambda) \sim \Delta(\lambda)^{\beta}\prod_{j=1}^N\lambda_j^{M-N}e^{-V(\lambda_j)}\int_{O(N), U(N), Sp(N)}e^{-tr(\Sigma^{-1}U\Lambda U^{-1})}dU,
%$$

%\ni where the integral for $\beta=1, 2, 4$, respectively, where $\Lambda = (\lambda_1, \dots, \lambda_N)$, $\Sigma = (l_1, \dots, l_N)$ is the covariance matrix. 
%\par In~\cite{KM99, KMNOY} the {\it generic} $\tau$-functions of Painlev\'e II and IV were shown to have Hankel determinant structure coming from the associated (one-dimensional) Toda equation. 
\par Generating functions of entries of generic Hankel determinant solutions to Painlev\'e II and IV were considered in~\cite{NaKaMaP2, NaKaMaP4}. They were shown to be simply related to an eigenvector component of the corresponding Lax pairs. E.g.~function $\F(t,x)$ of eq.~(4.22) with $\kappa=1$ appeared in~\cite{NaKaMaP2} where log-derivative of the first eigenvector component $Y_1$ of traceless Lax pair for Painlev\'e II~\cite{FN80, JM81} was identified as such a generating function. % and its connection with Toda chain was exposed 
In our normalization $\F(t,x)=F_2(t)\exp(-1/2(tx-x^3/3))Y_1$. It is an interesting open question if the Hankel determinant structure coming from Toda equations satisfied by Painlev\'e $\tau$-functions can be generalized to the other values of $\kappa=\beta/2$. %The works on one-dimensional Toda hierarchy and generating functions similar to $\F_{\kappa}$ for $\kappa=1$~\cite{NaKaMaP2, NaKaMaP4, KMO07} should be also helpful in this respect.
\par Remarkable expansions of generic $\tau$-functions for Painlev\'e VI, V and III around their regular (and irregular for Painlev\'e III) singular points with coefficients given by discrete Fourier transforms of Virasoro conformal blocks were conjectured in~\cite{GaIoLi, GIL653} using the combinatorial expansion of conformal blocks from the proof of AGT~\cite{AGT} correspondence by~\cite{AFLT} via representation theory of tensor product of Virasoro and Heisenberg algebras. The conformal expansions of Painlev\'e $\tau$-functions have the form

$$
\tau(t) = \sum_{n\in\mathbb Z} C(\{\theta_i\}; \sigma+n)s^n\F_{c=1}[\{\theta_i\}; \sigma+n](t),   \eqno(5.2)  %(5.3)  %(CBexp)
$$

\ni where $\theta_i$ are the monodromy exponents at the singular points $i$ (e.g.~$\theta_0, \theta_t, \theta_1, \theta_\infty$ for the standard form of Painlev\'e VI), the parameters $\sigma$ and $s$ correspond to Painlev\'e integration constants and $\F_{c=1}[\{\theta_i\}; \sigma+n](t)$ are $c=1$ CFT conformal blocks, e.g.~in case of Painlev\'e VI those related to the holomorphic part of 4-point correlator of the Virasoro primary operators located at $0, t, 1, \infty$. The parameters $\sigma$ and $s$ were studied by Jimbo in~\cite{Ji82} where the first terms of expansions (5.2) were found. Vastly extending his results, exact formulas for all the terms were obtained in~\cite{GaIoLi, GIL653}. This later led to finding exact formulas for the connection coefficients between the expansions of $\tau$-functions at different singular points for Painlev\'e VI~\cite{IoLiTy} and Painlev\'e III~\cite{ItsLiTy}.  
\par Matrix models for $\beta=2$ were considered from $c=1$ CFT point of view by~\cite{EyRi13} where it was shown that classical integrable structure e.g.~Lax matrix considerations may give more powerful results than just the application of general but complicated topological recursion of~\cite{EyEtAlTopBeta}. By showing how local conformal symmetry translates into isomonodromy of linear matrix first order ODE with Lax matrix of coefficients the authors~\cite{EyRi13} gave a clearer treatment to some of the results of~\cite{GaIoLi}. The classical integrable structure in the form of Lax pairs is exactly what was generalized in~\cite{betaFP1} to the other $\beta$.
\par Special solutions in terms of elliptic $\Theta$-functions for multitime $c=1$ BPZ equations were obtained in~\cite{DNov09}. This list of results could be continued -- many more exist for $\beta=2$ and everything is known in principle how to compute here.

\subsection{Results for $c\to\infty$ -- the quasiclassical (WKB) limit}

The general ideas and results are well described in a comprehensive review~\cite{Tesch10}. There are two different types of this limit: $\beta\to 0$ and $\beta\to\infty$. One of them is stationary, corresponding to the Nekrasov-Shatashvili (NS) limit~\cite{NS09, MMM11} in supersymmetric gauge theories and related to quantum integrable systems like quantum spin chains. Considering dependence of the limiting SE-type equations on their free accessory parameters leads to isomonodromy systems. On the other hand, using the links to quantum (gauge) theories and their NS limits can lead to determination of accessory parameters and solving the related uniformization of Riemann surfaces problem, see e.g.~\cite{NRS, Piat13} and references therein. A different stationary limit arises when one considers large matrix size $N$ limit together with $\alpha\sim1/N\to 0$ in the integrals of section 3~\cite{Bour12}. Then one obtains an SE of the form $(\hbar\prt)^2-V_{eff})\psi=0$ with $\hbar\sim(\beta/2-1)$~\cite{Bour12}, i.e.~there case $c\to1$ becomes the further classical limit. 
\par The other $c\to\infty$ limit leads to conventional classical integrable systems directly and is in this respect similar to the case $c=1 (\beta=2)$, but very different phenomena like Wigner crystallization of eigenvalue positions of $\beta$-ensembles~\cite{DE02, EdSut} also occur. Which of the limits corresponds to $\beta\to 0$ and which -- to $\beta\to\infty$, depends on the type of equation, (3.11) or (3.12). The stationary limit corresponds to strong diffusion, and the nonstationary -- to the weak one. In the last limit there are also recent new results on connection problem for Painlev\'e VI arising as the equation of motion from the classical action, the limit of the logarithm of conformal block~\cite{LLNZ}. This is one of the simplest cases of the problem of uniformization and finding the accessory parameters, which are on the one hand the derivatives of the classical Liouville theory action~\cite{TahZog} and on the other hand the values of the classical Gaudin Hamiltonians translating into the Garnier-type Hamiltonians~\cite{Tesch10}.
\par An important example of the nonstationary limit comes from the hydrodynamic (``collective field") description of the equations of motion of the classical CSM system (with arbitrary coupling constant $g$) for {\it any} number of particles including the infinite number of particles limit. It turns out~\cite{ABW09} to be equivalent to the complex (or {\it bidirectional}) Benjamin-Ono (BO) equation, i.e.~equation for a complex function $u$,

$$
\prt_tu + \prt_x\left(\frac{u^2}{2} + i\frac{g}{2}\prt_x\bar u\right) = 0,   \eqno(5.3)  %(5.4)  %(2BO)
$$

$$
u = u_0 + u_1, \qquad \bar u = u_0 - u_1,
$$

%$$
%u_1 = -i\frac{\pi g}{L} \lim_{N\to\infty} \sum_{j=1}^N \cot \frac{\pi}{L}(x-x_j),   \qquad   u_0 = i\frac{\pi g}{L} \lim_{N\to\infty} \sum_{j=1}^N \cot \frac{\pi}{L}(x-y_j),
%$$

$$
u_1 = -i\frac{\pi g}{L} \sum_{j=1}^N \cot \frac{\pi}{L}(x-x_j),   \qquad   u_0 = i\frac{\pi g}{L} \sum_{j=1}^N \cot \frac{\pi}{L}(x-y_j),
$$

\ni where $x_j$ are the coordinates of the CSM particles, $p_j$ are their momenta,

$$
\dot x_j(t) = p_j,   \qquad  \dot p_j(t) = -\left( \frac{\pi g}{L} \right)^2\prt_j\sum_{k\neq j}^N \left(\cot \frac{\pi}{L}(x_j-x_k) \right)^2,
$$

\ni  and the auxiliary complex variables $y_j$ are implicitly defined by the equations

$$
-\frac{2\pi}{L}p_j = i\frac{\dot w_j}{w_j} = \frac{g}{2}\left(\frac{2\pi}{L}\right)^2\left(\sum_{k=1}^N\frac{w_j+u_k}{w_j-u_k} - \sum_{k\neq j}^N\frac{w_j+w_k}{w_j-w_k}\right),
$$

\ni where $w_j = e^{2i\pi x_j/L}$ and $u_k = e^{2i\pi y_k/L}$. The coupling constant $g$ of the classical CSM system can be always rescaled to $1$ by rescaling $x$, $x_j$ and $t$. This is in contrast to the quantum CSM system considered in the similar way in~\cite{AbW05}, where the coupling constant is the physically important $\kappa=\beta/2$ parameter on which the properties of the system may crucially depend (and which cannot be removed by rescaling). 
\par If one splits the function $u$ into the real and imaginary parts,

$$
u = \mu + ir,
$$

\ni and introduces function $V=\prt_xr-r^2$, the complex BO equation can be rewritten as the real system

$$
\prt_tV + \mu\prt_xV + 2V\prt_x\mu + \frac{\prt_{xxx}\mu}{2} = 0,   \eqno(5.4)  %(5.5)  %(PBconn)
$$

$$
\prt_t\mu + \mathcal H\prt_{xx}\mu + \mu\prt_x\mu + \frac{\prt_xV}{2} = 0,   \eqno(5.5)  %(5.6)  %(C-law)
$$

\ni where the trigonometric Hilbert transform $\mathcal H$,

$$
(\mathcal H\varphi)(z) = v.p. \int_{|w|=1} \frac{dw}{2\pi iw} \varphi(w)\frac{e^{i(z-w)/2} + e^{-i(z-w)/2}}{e^{i(z-w)/2} - e^{-i(z-w)/2}}.   \eqno(5.6)  %(5.7)  %(HT)
$$

\ni  is used. Comparing now with our governing system for QPII in the form (4.38), (4.39), we see that the first equations are similar while the second ones are different though of the same one-dimensional hydrodynamics with pressure type. The universal equation (5.4) should be the same for all such systems, in geometry it expresses the relation between a Beltrami differential $\mu$ (recall our main governing function, the ratio of upper non-diagonal entries of Lax pair matrices) and a projective connection $V$ (Schr\"odinger potential, holomorphic part of conformal energy-momentum tensor), see e.g.~\cite{BFK91, NRS}. The last system of PDEs appeared in the description of quantum integrals of motion whose joint eigenfunctions make the appropriate orthogonal basis for combinatorial expansion of Liouville CFT conformal blocks~\cite{AFLT} important for establishing the AGT correspondence~\cite{AGT} with Nekrasov partition functions~\cite{NekOk} of supersymmetric gauge theory. Recently these considerations were also generalized~\cite{LitILW, QILW} to the quantum Intermediate Long Wave (ILW) hierarchy (related to more general versions of AGT correspondence), which continuously interpolates between two important limiting cases of (quantum) KdV and (quantum) BO hierarchies. The first is related to the studies of~\cite{BLZ, BLZ01, BLZ03}, e.g.~to the Schr\"odinger equation (4.1) considered above. % and thus should be related to QPIV at least in a certain special case
The governing system for the semiclassical limit of quantum ILW is similar to (5.4), (5.5)~\cite{LitILW, QILW}, differing only in the generalization of the Hilbert transform from trigonometric to elliptic version using theta-function $\Theta_1$,

%$$
%\prt_tV + v\prt_xV + 2V\prt_xv + \frac{\prt_{xxx}v}{2} = 0,   \eqno(PBconn)
%$$

%$$
%\prt_tv + \mathcal T\prt_{xx}v + v\prt_xv + \frac{\prt_xV}{2} = 0.   \eqno(C-law)
%$$

%\ni the only differerence being in the generalization of the Hilbert transform from the trigonometric to the elliptic version using theta-function $\Theta_1$,

$$
(\mathcal H f)(z) \to (\mathcal Tf)(z) = v.p. \int_{|w|=1}\frac{dw}{2\pi iw} f(w)\left(\ln \Theta_1((w-z)/2 | e^{-\tau})\right)',
$$

\ni with $\tau\to0$ and $\tau\to\infty$ corresponding to the KdV and the BO limits, respectively.
 
\section{Multidimensional FP operators}

%\subsection{Multi-Penner $\beta$-ensemble and correlation functions of CFT}

%A good starting point to demonstrate the connections between $\beta$-ensembles of RM theory (RMT), differential equations of CFT, Calogero-Sutherland-Moser models (CSM) and isomonodromy systems is the so-called multi-Penner $\beta$-ensemble~\cite{EgMar10, BoMarTan11, Bour12} defined by eigenvalue integral
The following integral defines a multi-Penner $\beta$-ensemble and also, up to a simple factor, certain Liouville CFT correlation function~\cite{BoMarTan11, AgEtAl11, Bour12}:

%$$
%Z_N = \prod_{i<j}(z_i-z_j)^{-1/2b^2}\prod_{1\le k<j\le n}(w_k-w_j)^{-2m_km_j/g_s^2}\prod_{i=1}^l\prod_{k=1}^{n}(z_i-w_k)^{-m_k/bg_s}\cdot
%$$

%$$
%\cdot \int\dots\int \prod_{i=1}^Ndx_i\prod_{i<j}(x_i-x_j)^{-2b^2}\prod_{i=1}^N\prod_{k=1}^{n}(x_i-w_k)^{2bm_k/g_s}\prod_{j=1}^l(z_j-x_i),   \eqno(6.1)  %(ZmP)
%$$

$$
Z_N = \int\dots\int \prod_{i=1}^Ndx_i\prod_{i<j}(x_i-x_j)^{\beta}\prod_{i=1}^N\prod_{k=1}^{n}(x_i-w_k)^{-C(N,\beta)m_k}\prod_{j=1}^l\prod_{i=1}^N(z_j-x_i),   \eqno(6.1)  %(ZmP)
$$

\ni In Liouville CFT~~\cite{BoMarTan11, AgEtAl11, Bour12} it corresponds to a $\beta$-ensemble with the multi-Penner potential (3.9) with $C = -2b/g_s$, where $g_s$ is the gauge or string coupling constant, but there $\beta=-2b^2<0$\footnote{so the integrations should be performed over a contour in complex plane where the integrals converge} and central charge $c>25$, see (2.2). The additional mass parameter $m_{n+1}$ of the theory implied to be at $\infty$ in CFT correlators satisfies the constraint

$$
\frac{b}{g_s}\sum_{k=1}^{n+1}m_k + \frac{l}{2} = b^2N - b^2 - 1.
$$

% Clearly here shift in the semiinfinite integration domain leads to the same integral with just correspondingly shifted $\{z_i\}$ and $\{w_k\}$ parameters.
\ni Integrals of this type are important in the AGT correspondence~\cite{AGT}, see e.g.~\cite{BoMarTan11}. Such an integral satisfies the linear partial differential equations (PDEs) derived as in section 3,

%$$
%\left(b^2\prt_{z_iz_i} + \sum_{j\neq i}^l\left(\frac{1}{z_i-z_j}\prt_{z_j} - \frac{3+2b^2}{4b^2(z_i-z_j)^2}\right) + \sum_{k=1}^n\left(\frac{1}{z_i-w_k}\prt_{w_k} + \frac{\Delta_k}{(z_i-w_k)^2}\right) \right)Z_N = 0   \eqno(6.2)  %(T-BPZ)    %(Tesch-5.24)
%$$

$$
\left(\frac{\beta}{2}\prt_{z_iz_i} + \sum_{j\neq i}^l\frac{1}{z_i-z_j}(\prt_{z_i}-\prt_{z_j}) - C(N,\beta)\sum_{k=1}^n\frac{m_k}{z_i-w_k}\prt_{z_i} + \sum_{k=1}^n\frac{1}{z_i-w_k}\prt_{w_k} \right)Z_N = 0   \eqno(6.2)  %(T-BPZ)    %(Tesch-5.24)
$$

%\ni where $\Delta_k = b^2m_k^2/g_s^2+(b+1/b)^2/4$ are the conformal (scaling) dimensions. 
\ni These are the more general BPZ equations -- the equations satisfied by correlation functions of CFT with $l$ ``degenerate at level two" primary fields (operator generating functions) at $z_i, i=1,\dots,l$ included~\cite{BPZ}. This fact was implicitly used immediately after the creation of CFT~\cite{BPZ} in the Coulomb gas approach~\cite{DF}, where more general multidimensional integrals over complex eigenvalues appeared, even before the Virasoro constraints were introduced as a main tool to tackle matrix integrals related to gauge QFT and lower-dimensional quantum gravity. %We see the appearance of another Calogero interaction among the variables $z_i$. 
\par On the one hand, studies of the soft edge limit of $\beta$-ensemble with external matrix source with $r$ non-zero eigenvalues ($r$ {\it spikes} in statistics community terminology) by probabilistic methods yielded~\cite{BV2} the multidimensional FP equation corresponding to QPII for $r=1$,

$$
\left[r\prt_t + \sum_{k=1}^r\left(\frac{2}{\beta}\frac{\prt^2}{\prt x_k^2} + (t-x_k^2)\frac{\prt}{\prt x_k} + \sum_{j\neq k}^r\frac{1}{x_k-x_j}\left(\frac{\prt}{\prt x_k} - \frac{\prt}{\prt x_j}\right) \right)\right]\F(t,x) = 0.   \eqno(6.3)  %(BV2)
$$

\ni This implies the presence of $r$ degenerate level two primaries in the corresponding CFT correlation function. We recall that in general an ODE from CFT has first order w.r.t.~the locations of non-degenerate primary operators and second order w.r.t.~the degenerate level 2 primaries. Up to a change $\beta/2\to 2/\beta$ in (6.2), the PDE (6.3) can be obtained as the sum from $1$ to $r$ of triconfluent limits of equations (6.2) with $n=3$ and $l=r$ leading to just one ``time" variable (the cross-ratio of the $w_k$ including $w_4=\infty$). Details will appear elsewhere.
\par A multidimensional PDE of the type (6.3) but nonconfluent (i.e.~generalizing QPVI rather than QPII) and in elliptic form like (4.45) was considered by~\cite{LanTak12} and elliptic solutions for special values of $s_k$-like parameters (see (4.45)) were obtained far extending the result of~\cite{FLNO}. It would be interesting to find all possible confluent limits of these solutions. 
\par A linear PDE of different but related type -- multidimensional ``quantum isomonodromy" equation with one ``time" and $N$ Garnier-like coordinates as independent variables --was proposed by Yamada~\cite{Ya2011} to describe the instanton partition function of superconformal gauge theory with symmetry group $SU(N)$. %His equation corresponds to quantum Garnier type systems coupled in a certain way.
\par Returning to CSM systems, if one introduces in (2.12) the power-sum operators

$$
a_n = \frac{\prt}{\prt t_n},  \qquad a_{-n} = a^\dagger_n = \frac{nt_n}{\kappa}, \quad n\ge0  %\eqno(2.13)
$$

\ni  in terms of $t_n=\sum_jy_j^k/k$, satisfying the Heisenberg algebra $[a_n, a^\dagger_m] = \frac{n}{\kappa}\delta_{n,m}, \quad (n, m > 0)$, one can rewrite the modified CSM Hamiltonian (2.12) as

%$$
%[a_n, a^\dagger_m] = \frac{1}{\kappa}n\delta_{n,m}, \quad (n, m > 0)  \eqno(2.14)
%$$

%\ni and acting on vacuum states of the quantum CSM so that $a_n|0\rangle = \langle0|a^\dagger_n = 0$, $n>0$, and taking the generating function $A(y) = \sum_{n=1}^\infty\frac{a_n}{n}y^n$,  one can translate the collective field Hilbert space into the space of symmetric polynomials in $y$-variables by the formula

%$$
%\langle0|e^{\kappa\sum_{i=1}^NA(y_i)}a^\dagger_{n_1}\dots a^\dagger_{n_m}|0\rangle = p_{n_1}\dots p_{n_m} = n_1\dots n_mt_1\dots t_m.   \eqno(2.15)
%$$

%\ni This convention reproduces the standard inner product between symmetric polynomials~\cite{Mac}. The operator $\tilde H$ satisfies the intertwining property

%$$
%\tilde H \langle0|e^{\kappa\sum_{i=1}^NA(y_i)} = \langle0|e^{\kappa\sum_{i=1}^NA(y_i)} \hat H,   \eqno(2.16)
%$$

%\ni where

%$$
%\hat H = \kappa^2\sum_{n,m=1}^\infty(a^\dagger_{n+m}a_na_m + a^\dagger_na^\dagger_ma_{n+m}) + \kappa(1-\kappa)\sum_{n=1}^\infty na^\dagger_na_n + \kappa^2N\sum_{n=1}^\infty a^\dagger_na_n,   \eqno(2.17)
%$$

%$$
%\tilde H = \kappa^2\sum_{n,m=1}^\infty(a^\dagger_{n+m}a_na_m + a^\dagger_na^\dagger_ma_{n+m}) + \kappa(1-\kappa)\sum_{n=1}^\infty na^\dagger_na_n + \kappa^2N\sum_{n=1}^\infty a^\dagger_na_n,   \eqno(2.17)
%$$

%\ni which can also be written as

%$$
%\hat H = \kappa\sum_{n=1}^\infty a^\dagger_nL_n + (\kappa(N+1-2a_0) - 1)\kappa\sum_{n=1}^\infty a^\dagger_na_n.   \eqno(2.17a)
%$$

$$
\tilde H_{CSM} = \kappa\sum_{i,j\ge1}\left( \kappa ijt_{i+j}\frac{\prt^2}{\prt t_i\prt t_j} + (i+j)t_it_j\frac{\prt}{\prt t_{i+j}} \right) + \kappa(\kappa-1)\sum_{i\ge1}i^2t_i\frac{\prt}{\prt t_i}.  \eqno(6.4)
$$

\ni Thus, the CSM diffusion-drift operator for infinite number of particles expressed in terms of the power sums of the particle coordinates becomes the $\kappa$-deformed {\it cut-and-join operator}~\cite{GouldJack}. The cut-and-join operator corresponding to $\kappa=1$ is important in enumerative geometry and representation theory of the symmetric group. In this case, certain exponential generating function of Hurwitz numbers was shown~\cite{Kaz09} to satisfy the infinite-dimensional heat-type equation

%$$
%\kappa\sum_{i,j\ge1}\left( \kappa ijt_{i+j}\frac{\prt^2}{\prt t_i\prt t_j} + (i+j)t_it_j\frac{\prt}{\prt t_{i+j}} \right) + \kappa(\kappa-1)\kappa\sum_{i\ge1}i^2t_i\frac{\prt}{\prt t_i}.
%$$

%\ni Again the Goulden operator corresponds to $\kappa=1$. For this case, a special exponential generating function of Hurwitz numbers was shown~\cite{Kaz09} to satisfy the infinite-dimensional heat-type equation

$$
\frac{\prt e^{H(s, {\bf t})}}{\prt s} = \frac{1}{2}\sum_{i,j\ge1}\left( ijt_{i+j}\frac{\prt^2}{\prt t_i\prt t_j} + (i+j)t_it_j\frac{\prt}{\prt t_{i+j}} \right)e^{H(s, {\bf t})},   \eqno(6.5)  %(hCSM1)
$$

\ni which makes its solution $e^{H(s, {\bf t})}$ a $\tau$-function of the KP hierarchy. Different diffusion operators of this type appear in the theory of so-called z-measures~\cite{BorOl} in representation theory of the symmetric group.

%$$
%$$

%\ni ?: Comment on separation of variables and reduction to 1d diffusions.
\par Using the operator for general $\kappa$, the quantum CSM Hamiltonian, one can derive the correspondence between CSM with infinite number of particles and quantized Benjamin-Ono (QBO) equation~\cite{AbW05, NazSkl12},

$$
\tilde H_{CSM} = \frac{\kappa^3}{3}\sum_{l+m+n=0}: a_la_ma_n :  + \frac{\kappa^2(\kappa-1)}{2}\sum_{m+n=0}|n|: a_ma_n :  = 
$$

$$
= \int_{S^1}\frac{dz}{2\pi iz}\left(\frac{1}{3}:\varphi^3(z): + \frac{\kappa-1}{2}iz:\varphi'(z)(\mathcal H\varphi)(z):\right) = H_{QBO},   \eqno(6.6)  %(HBO)
$$

%\ni where the generating operator function $\phi(z)$ (denoted $\prt\phi(z) in (2.7))

%$$
%\phi(z) = \sum_{n\neq0}\frac{a_n}{z^n}    \eqno(phi)
%$$

\ni where the generating operator function $\varphi(z)$ is the collective field variable denoted $\prt\phi(z)$ in (2.6), and $\mathcal H\varphi$ is the Hilbert transform defined by (5.6). In~\cite{NazSkl13, NazSkl12} exact generating function of an infinite family of commuting integrals of the quantum BO equation is derived using its quantum Lax matrix in the power-sum representation. %(which has a suggestive Toeplitz structure). 
\par It seems natural to raise the question of generalization of the classical ILW governing system to the quantum case for any $\kappa=\beta/2$ which would give an exact quantum-classical correspondence similar to the one we exposed for the QPII in section 4.2. We conjecture that such a {\it classical} governing system exists, and its first equation is similar to (5.4), up to the change $\prt_t\to\hbar\prt_t$ as in (4.38) but the second equation -- substitute for (5.5) -- may change more substantially. Interesting hints about what happens can be found in~\cite{AbW05}, e.g.~here likely $\hbar\sim2/(\sqrt\kappa-1/\sqrt\kappa)$, but more study is needed.

%\section{Genuine quantum systems and quantum-classical correspondence}
%These are quantum Calogero-Sutherland-Moser systems, non-commutative quantum integrable systems such as quantum Gaudin models, quantum spin chains (XXX, XXZ, XYZ chains and their degenerations).

%\section{SLE processes, quantum Hall effect, 2d CFT}
  
\section{Concluding remarks} 

Quantum Painlev\'e equations and their multidimensional generalizations are equivalent to the (confluent) BPZ equations~\cite{BPZ} of CFT. They are satisfied by averaged powers of characteristic polynomials of general $\beta$-ensembles of random matrices which are excellent toy models to study various properties of CFT. These linear PDEs themselves carry all information about associated isomonodromic nonlinear integrable systems. Such equations establish an exact quantum-classical correspondence by means of which quantum integrable systems find their equivalent description in terms of classical integrable isomonodromy systems. Thus, these special ``quantum conformal" PDEs deserve the most thorough analytic investigation.
\par We do not consider it but everything here should readily generalize to the case of different symmetry group (here we implicitly considered $GL(2)$) where $W$-algebras substitute the Virasoro algebra and PDEs of higher order arise, and to the $q$-deformed models where Jack functions and CSM operators are replaced by Macdonald functions and difference operators, see e.g.~\cite{BC-Mac}, and so the $q$-difference equations should naturally arise instead of the considered PDEs.
\par A number of topics, without which the picture drawn here is very incomplete, remained out of the scope of this paper. They include the relations between BPZ~\cite{BPZ} and KZ~\cite{KZ} equations of CFT, quantum spin chains and their recently discovered connection with classical integrable hierarchies~\cite{AKLTZ}, the use of Nekrasov functions~\cite{NekOk} (the other side of AGT correspondence), complex $\beta$-ensembles, Stochastic L\"owner Evolutions (SLE) and their integrable description coming from that of CFT and $\beta$-ensembles, the recently emerged field of integrable probability and Macdonald processes~\cite{BC-Mac}, non-commutative integrable systems, see e.g.~\cite{JimNagSu08}, and their possible description in terms of commutative ones. At last, the unifying link should be provided by the discrete Hirota bilinear equations reviewed in~\cite{TYsys}, hints of which appeared e.g.~in~\cite{KLWZ97}, and in~\cite{ODEIM} and references therein, see section 4.1, and in~\cite{AKLTZ}. We plan to address these topics in the second part of this review, in progress.

\bigskip

{\bf\large Acknowledgments.} The author is grateful to A.~Dzhamay, K.~Maruno and C.~Ormerod for the invitation to write this review for JMM 2014 Proceedings, to M.~Ablowitz and R.~Maier for useful discussions. Partial support by NSF grant DMS-0905779 is acknowledged.


\bigskip
\begin{thebibliography}{10}

{\small

\bibitem{AbW05}
A.~Abanov, P.~Wiegmann.
\newblock Quantum hydrodynamics, quantum Benjamin-Ono equation, and Calogero model.
\newblock {\em Phys. Rev. Lett. }, 95:076402, 2005; {\em arXiv:cond-mat/0504041}.  

\bibitem{ABW09}
A.~Abanov, E.~Bettelheim, P.~Wiegmann.
\newblock Integrable hydrodynamics of Calogero-Sutherland model: bidirectional Benjamin-Ono equation.
\newblock {\em J. Phys. A}, 42:135201, 2009; {\em arXiv:0810.5327}.  

\bibitem{AbF2003}
M.~Ablowitz, A.~Fokas.
\newblock Complex Variables.
\newblock {\em 2nd edition, Cambridge University Press, Cambridge, UK}, 2003.  

\bibitem{AgEtAl11}
M.~Aganagic, M.~Cheng, R.~Dijkgraaf, D.~Krefl, C.~Vafa.
\newblock Quantum geometry of refined topological strings.
\newblock {\em J. High Energy Phys.}, 11:019, 2012; {\em arXiv:1105.0630}.

\bibitem{AFLT}
V.~Alba, V.~Fateev, A.~Litvinov, G.~Tarnopolskiy.
\newblock On combinatorial expansion of the conformal blocks arising from AGT conjecture.
\newblock {\em Lett. Math. Phys.}, 98:33--64, 2011; {\em arXiv:1012.1312}.

\bibitem{AGT}
L.~Alday, D.~Gaiotto, Y.~Tachikawa.
\newblock Liouville correlation functions from four-dimensional gauge theories.
\newblock {\em Lett. Math. Phys.}, 91:167--197, 2010; {\em arXiv:0906.3219v2}.

\bibitem{AKLTZ}
A.~Alexandrov, V.~kazakov, S.~Leurent, Z.~Tsuboi, A.~Zabrodin.
\newblock Classical $\tau$-function for quantum spin chains.
\newblock {\em J. High Energy Phys.}, 1309:064, 2013; {\em arXiv:1112.3310v2}.  
A.~Alexandrov, S.~Leurent, Z.~Tsuboi, A.~Zabrodin.
\newblock The master $T$-operator for the Gaudin model and the KP hierarchy.
\newblock {\em Nucl. Phys. B}, 883:173--223, 2014; {\em arXiv:1306.1111}.  

\bibitem{ABG12}
R.~Allez, J.-P. Bouchaud, A.~Guionnet.
\newblock Invariant $\beta$-ensembles and the Gauss-Wigner crossover.
\newblock {\em Phys. Rev. Lett.}, 109, id:094102, 2012; {\em arXiv:1205.3598}

%\bibitem{ArEtAl2010}
%D.~Arrigo, D.~Ekrut, J.~Fliss, L.~Le.
%\newblock Nonclassical symmetries of a class of Burgers' systems.
%\newblock {\em J. Math. Anal. Appl.}, 371:813--820, 2010.

\bibitem{AwMOSh}
H.~Awata, Y.~Matsuo, S.~Odake, J.~Shiraishi.
\newblock Collective Field Theory, Calogero-Sutherland Model and Generalized Matrix Models.
\newblock {\em Phys. Lett. B}, 347:49, 1995; {\em arXiv:hep-th/9411053v3}.

\bibitem{AwEtAl10}
H.~Awata, H.~Fuji, H.~Kanno, M.~Manabe, Y.~Yamada.
\newblock Localization with a surface operator, irregular conformal blocks and open topological string.
\newblock {\em Adv. Theor. Math. Phys.}, 16:725--804, 2012; {\em arXiv:1008.0574v5}.

\bibitem{BaikP06}
J.~Baik.
\newblock Painlev\'e formulas of the limiting distributions for nonnull complex sample covariance matrices.
\newblock {\em Duke Math. J.}, 133:205--235, 2006; {\em arXiv:math/0504606}.

\bibitem{BBP}
J.~Baik, G. Ben Arous, S.~Pech\'e.
\newblock Phase transition of the largest eigenvalue for nonnull complex sample covariance matrices.
\newblock {\em Ann. Prob.}, 33:1643--1697, 2005; {\em arXiv:math/0403022}.

\bibitem{BR01}
J.~Baik, E.~Rains.
\newblock Limiting distributions for a polynuclear growth model with external sources.
\newblock {\em J. Stat. Phys.}, 100:523--541, 2000; {\em arXiv:math/0003130}.

\bibitem{BLZ}
V.~Bazhanov, S.~Lukyanov, A.~Zamolodchikov.
\newblock Integrable Structure of Conformal Field Theory I, II, III.
\newblock {\em Comm. Math. Phys.}, 177:381--398, 1996; 190:247--278, 1997; 200:297--324, 1999; {\em arXiv:hep-th/9412229}; {\em arXiv:hep-th/9604044}; {\em arXiv:hep-th/9805008}.

\bibitem{BLZ01}
V.~Bazhanov, S.~Lukyanov, A.~Zamolodchikov.
\newblock Spectral determinants for Schr\"odinger equation and Q-operators of Conformal Field Theory.
\newblock {\em J. Stat. Phys.}, 102:567, 2001; {\em arXiv: hep-th/9812247v2}.

\bibitem{BLZ03}
V.~Bazhanov, S.~Lukyanov, A.~Zamolodchikov.
\newblock Higher level eigenvalues of Q-operators and Schr\"odinger equation.
\newblock {\em Adv. Theor. Math. Phys.}, 7:711--725, 2003; {\em arXiv: hep-th/0307108v2}.

\bibitem{BM05}
V.~Bazhanov, V.~Mangazeev.
\newblock Eight-vertex model and non-stationary Lam\'e equation.
\newblock {\em J. Phys. A}, 38:L145--L153, 2005; {\em arXiv: hep-th/0411094}.

\bibitem{BM0610}
V.~Bazhanov, V.~Mangazeev.
\newblock The eight-vertex model and Painlev\'e VI.
\newblock {\em J. Phys. A}, 39:12235--12243, 2006; {\em arXiv: hep-th/0602122}.
\newblock The eight-vertex model and Painlev\'e VI equation II: eigenvector results.
\newblock {\em J. Phys. A}, 43:085206, 2010; {\em arXiv: 0912.2163}.

\bibitem{BPZ}
A.~Belavin, A.~Polyakov, A.~Zamolodchikov.
\newblock Infinite conformal symmetry in two-dimensional quantum field theory.
\newblock {\em Nulc. Phys. B}, 241:333--380, 1984.

\bibitem{BerCaf14}
M.~Bertola, M.~Cafasso.
\newblock Darboux transformations and random point processes.
\newblock {\em arXiv:1401.4752}, 2014.

\bibitem{BFK91}
A.~Bilal, V.~Fock, I.~Kogan.
\newblock On the origin of W-algebras.
\newblock {\em Nulc. Phys. B}, 359:635--672, 1991.

\bibitem{BV1}
A.~Bloemendal, B.~Virag.
\newblock Limits of spiked random matrices I.
\newblock {\em Theor. Prob. Rel. Fields}, 156:795--825, 2013; {\em arXiv:1011.1877}.

\bibitem{BV2}
A.~Bloemendal, B.~Virag.
\newblock Limits of spiked random matrices II.
\newblock {\em arXiv:1109.3704}, 2011.

\bibitem{BC69}
G.~Bluman, J.~Cole.
\newblock The general similarity solution of the heat equation.
\newblock {\em J. Math. Mech.}, 18:1025--1042, 1969.

\bibitem{BoMarTan11}
G.~Bonelli, K.~Maruyoshi, A.~Tanzini.
\newblock Quantum Hitchin systems via $\beta$-deformed matrix models.
\newblock {\em arXiv:1104.4016v2}.  

\bibitem{QILW}
G.~Bonelli, A.~Sciarappa, A.~Tanzini, P.~Vasko.
\newblock Six-dimensional supersymmetric gauge theories, quantum cohomology of instanton moduli spaces and $gl(N)$ Quantum Intermediate Long Wave Hydrodynamics.
\newblock {\em arXiv:1306.0659}, 2013.  

\bibitem{BC-Mac}
A.~Borodin, I.~Corwin.
\newblock Macdonald processes.
\newblock {\em Prob. Th. Rel. Fields}, 158:225--400, 2014; {\em arXiv:1111.4408}, 2011.  \\
A.~Borodin, I.~Corwin, V.~Gorin, Sh.~Shakirov.
\newblock Observables of Macdonald processes.
\newblock {\em arXiv:1306.0659}, 2013. 

%\bibitem{BCGSh}
%A.~Borodin, I.~Corwin, V.~Gorin, Sh.~Shakirov.
%\newblock Observables of Macdonald processes.
%\newblock {\em arXiv:1306.0659}, 2013.  

\bibitem{BorOl}
A.~Borodin, G.~Olshanski.
\newblock Infinite-dimensional diffusions as limits of random walks on partitions.
\newblock {\em Prob. Th. Rel. Fields}, 144:281--319, 2009 ; {\em arXiv: 0706.1034}.  \\
G.~Olshanski.
\newblock Anisotropic Young diagrams and infinite-dimensional diffusion processes with Jack parameter.
\newblock {\em Intern. Math. Res. Notices}, rnp168:1102--1166, 2010 ; {\em arXiv: 0902.3395}.  

\bibitem{Bour12}
J.~E.~Bourgine.
\newblock Large $N$ limit of $\beta$-ensembles and deformed Seiberg-Witten relations.
\newblock {\em J. High Energy Phys.}, 08:46, 2012; {\em arXiv:1206.1696v2}.

\bibitem{EyEtAlTopBeta}
B.~Eynard, N.~Orantin.
\newblock Topological expansion of the Bethe ansatz and quantum algebraic geometry.
\newblock {\em Commun. Number Th.}, 1:347--452, 2007; {\em arXiv:math-ph/0702045v4}.  \\
L.~Chekhov, B.~Eynard, O.~Marchal.
\newblock Topological expansion of the Bethe ansatz and quantum algebraic geometry.
\newblock {\em arXiv:0911.1664}.

\bibitem{Der09}
P.~Desrosiers.
\newblock Dualities at all beta.
\newblock {\em Nucl. Phys. B}, 36:2963--2981, 2009; {\em arXiv:0801.3438}.

\bibitem{DerLiu13}
P.~Desrosiers, D.-Z.~Liu.
\newblock Scaling limits of correlations of characteristic polynomials for the Gaussian $\beta$-ensemble with external source.
\newblock {\em Int. Math. Res. Not.}, rnu039:1--31, 2014; {\em arXiv:1306.4058v3}.

\bibitem{ODEIM}
P.~Dorey, C.~Dunning, R.~Tateo.
\newblock The ODE/IM correspondence.
\newblock {\em J. Phys. A}, 40:R205--R283, 2007; {\em arXiv:hep-th/0703066}.

\bibitem{DorTat99}
P.~Dorey, R.~Tateo.
\newblock Anharmonic oscillators, the thermodynamic Bethe ansatz and nonlinear integral equations.
\newblock {\em J. Phys. A}, 32:L419, 1999; {\em arXiv:hep-th/9812211}.

\bibitem{DTCFT}
P.~Dorey, R.~Tateo.
\newblock On the relation between Stokes multipliers and the T-Q systems of conformal field theory.
\newblock {\em Nucl. Phys. B}, 563:573--602, 1999; {\em arXiv:hep-th/9906219}.

\bibitem{DF}
Vl.~Dotsenko, V.~Fateev.
\newblock Conformal algebra and multipoint correlation functions in 2D statistical models.
\newblock {\em Nucl. Phys. B}, 240:312--348, 1984.
\newblock Four-point correlation functions and the operator algebra in 2D conformal invariant theories with central charge $c\le1$.
\newblock {\em Nucl. Phys. B}, 251[FS13]:691--734, 1985.

\bibitem{EdEtAlBeta13}
A.~Dubbs, A.~Edelman, P.~Koev, P.~Venkataramana.
\newblock The Beta-Wishart ensemble.
\newblock {\em arXiv:1305.3561}, 2013.  

\bibitem{DubMa07}
B.~Dubrovin, M.~Mazzocco.
\newblock Canonical structure and symmetries of the Schlesinger equations.
\newblock {\em J. Math. Phys.}, 271:289--373, 2007; {\em arXiv:math/0311261v4}.

\bibitem{DE02}
I.~Dumitriu, A.~Edelman.
\newblock Matrix models for beta ensembles.
\newblock {\em J. Math. Phys.}, 43:5830--5847, 2002; {\em arXiv:math-ph/0206043}.

\bibitem{DyBeta}
F.~Dyson.
\newblock A Brownian motion model for the eigenvalues of a random matrix.
\newblock {\em J. Math. Phys.}, 3:1191--1198, 1962.

\bibitem{EdSut}
A.~Edelman, B.~Sutton.
\newblock From random matrices to stochastic operators.
\newblock {\em J. Stat. Phys.}, 127:1121--1165, 2007; {\em arXiv:math-ph/0607038}.

\bibitem{EtKir93}
P.~Etingof, A.~Kirillov~Jr.
\newblock Representations of affine Lie algebras, parabolic differential equations and Lam\'e functions.
\newblock {\em Duke Math. J.}, 74:585--614, 1994; {\em arXiv:hep-th/9310083v2}.

\bibitem{EyRi13}
B.~Eynard, S.~Ribault.
\newblock Lax matrix solution of $c=1$ Conformal Field Theory.
\newblock {\em arXiv:1307.4865v2}.

\bibitem{FLNO}
V.~Fateev, A.~Litvinov, A.~Neveu, E.~Onofri.
\newblock A differential equation for a four-point correlation function in Liouville field theory and elliptic four-point conformal blocks.
\newblock {\em J. Phys. A}, 42:304011, 2009; {\em arXiv:0902.1331}.

\bibitem{Fer13}
P.~Ferrari.
\newblock Why random matrices share universal processes with interacting particle systems?
\newblock {\em arXiv:1312.1126v2}, 2013.

\bibitem{FN80}
H.~Flaschka, A.~Newell.
\newblock Monodromy and spectrum preserving deformations.
\newblock {\em Comm. Math. Phys.}, 76;65--116, 1980.

\bibitem{FIKN}
A.~Fokas, A.~Its, A.~Kapaev, V.~Novokshenov.
\newblock Painlev\'e transcendents: the Riemann-Hilbert approach.
\newblock {\em American Mathematical Society, Providence, RI}, 2006.

\bibitem{F2010}
P.~Forrester.
\newblock Log-gases and Random Matrices.
\newblock {\em Princeton University Press, Princeton, NJ}, 2010.

%\bibitem{ForEvenBeta09}
%P.~Forrester.
%\newblock A Random Matrix Decimation Procedure Relating $\beta = 2/(r + 1)$ to $\beta = 2(r + 1)$.
%\newblock {\em Commun. Math. Phys.}, 285:653--672, 2009.

\bibitem{ForBeta11}
P.~Forrester.
\newblock Probability densities and distributions for spiked Wishart $\beta$-ensembles.
\newblock {\em arXiv:1101.2261v2}, 2011.

\bibitem{Gai}
D.~Gaiotto.
\newblock Asymptotically free $\mathcal N = 2$ theories and irregular conformal blocks.
\newblock {\em arXiv:0908.0307}.

\bibitem{GaiTesch}
D.~Gaiotto, J.~Teschner.
\newblock Irregular singularities in Liouville theory and Argyres-Douglas type gauge theories.
\newblock {\em J. High Energy Phys.}, 12:050, 2012; {\em arXiv:1203.1052}.

\bibitem{GaIoLi}
O.~Gamayun, N.~Iorgov, O.~Lisovyy.
\newblock Conformal field theory of Painlev\'e VI. 
\newblock {\em J. High Energy Phys.}, 10:038, 2012; {\em arXiv:1207.0787}, 2012.

\bibitem{GIL653}
O.~Gamayun, N.~Iorgov, O.~Lisovyy.
\newblock How instanton combinatorics solves Painlev\'e VI, V and IIIs.
\newblock {\em J. Phys. A}, 46:335203, 2013; {\em arXiv:1302.1832}, 2013.

\bibitem{GiVes09}
J.~Gibbons, A.~Veselov.
\newblock On the rational monodromy-free potentials with sextic growth.
\newblock {\em J. Math. Phys.}, 50:013513, 2009; {\em arXiv:0807.3501}.

\bibitem{GouldJack}
I.~Goulden.
\newblock A differential operator for symmetric functions and the combinatorics of multiplying transpositions.
\newblock {\em Trans. Amer. Math. Soc.}, 344:421--440, 1994.  \\
I.~Goulden, D.~Jackson.
\newblock Transitive factorizations into transpositions and holomorphic mappings on the sphere.
\newblock {\em Proc. Amer. Math. Soc.}, 125:51--60, 1997.

%\bibitem{HMcL80}
%S.~Hastings, J.~McLeod.
%\newblock A boundary value problem associated with the second Painlev\'e transcendent and the Korteweg-de Vries equation.
%\newblock {\em Arch. Rat. Mech. Anal.}, 73:31--51, 1981.

%\bibitem{Hir81}
%R.~Hirota.
%\newblock Discrete analog of a generalized Toda equation.
%\newblock {\em J. Phys. Soc. Japan.}, 50:3785, 1981.

\bibitem{IoLiTy}
N.~Iorgov, O.~Lisovyy, Yu.~Tykhyy.
\newblock Painlev\'e VI connection problem and monodromy of $c=1$ conformal blocks. 
\newblock {\em J. High Energy Phys.} 12:029, 2013; {\em arXiv:1308.4092v2}, 2013.

\bibitem{ItsLiTy}
A.~Its, O.~Lisovyy, Yu.~Tykhyy.
\newblock Connection problem for the sine-Gordon/Painlev\'e III tau function and irregular conformal blocks. 
\newblock {\em arXiv:1403.1235}, 2014.

\bibitem{GaussToP}
K.~Iwasaki, H.~Kimura, S.~Shimomura, M.~Yoshida.
\newblock From Gauss to Painlev\'e: a modern theory of special functions.
\newblock {\em Viewveg, Braunschweig, Germany}, 1991.

\bibitem{JM81}
M.~Jimbo, T.~Miwa.
\newblock Monodromy preserving deformations of linear ordinary differential equations II.
\newblock {\em Physica D2}, 407, 1981.

\bibitem{Ji82}
M.~Jimbo.
\newblock Monodromy problem and the boundary condition for some Painlev\'e equations.
\newblock {\em Publ. RIMS Kyoto Univ.}, 18:1137--1161, 1982.

\bibitem{JimNagSu08}
M.~Jimbo, H.~Nagoya, J.~Sun.
\newblock Remarks on the confluent KZ equation for $sl_2$ and quantum Painlev\'e equations.
\newblock {\em J. Phys. A}, 41:175205, 2008.

\bibitem{NaKaMaP2}
N.~Joshi, K.~Kajiwara, M.~Mazzocco.
\newblock Generating function associated with the determinant formula for solutions of the Painlev\'e II equation.
\newblock {\em Ast\'erisque}, 274:67--78, 2004; {\em arXiv:nlin/0406035}.

\bibitem{NaKaMaP4}
N.~Joshi, K.~Kajiwara, M.~Mazzocco.
\newblock Generating function associated with the determinant formula for solutions of the Painlev\'e IV equation.
\newblock {\em Funkcial. Ekvac.}, 49:451--468, 2006; {\em arXiv:nlin/0512041}.

%\bibitem{KMO07}
%K.~Kajiwara, M.~Mazzocco, Y.~Ohta.
%\newblock A remark on the Hankel determinant formula for solutions of the Toda equation.
%\newblock {\em J. Phys. A}, 40:12661--12675, 2007.

\bibitem{Kaw03}
H.~Kawamuko.
\newblock On holonomic deformation of linear differential equations with a regular singular point and an irregular singular point.
\newblock {\em Kyushu J. Math.}, 57:1--28, 2003.

\bibitem{Kaz09}
M.~Kazarian.
\newblock KP hierarchy for Hodge integrals.
\newblock {\em Adv. Math.}, 221:1--21, 2009; {\em arXiv:math/0809.3263}.

\bibitem{Kim89}
H.~Kimura.
\newblock The degeneration of the two-dimensional Garnier system and the polynomial Hamiltonian structure.
\newblock {\em Annali di Mat. pur. appl.}, CLV:25--74, 1989.

\bibitem{KZ}
V.~Knizhnik, A.~Zamolodchikov.
\newblock Current algebra and Wess-Zumino model in two dimensions.
\newblock {\em Nucl. Phys. B}, 247:83--103, 1984.

\bibitem{Kostov99}
I.~Kostov.
\newblock Conformal field theory techniques for matrix models.
\newblock {\em arXiv:hep-th/9907060}.

\bibitem{KLWZ97}
I.~Krichever, O.~Lipan, P.~Wiegmann, A.~Zabrodin.
\newblock Quantum Integrable Models and Discrete Classical Hirota Equations.
\newblock {\em Comm. Math. Phys.}, 188:267--304, 1997; {\em arXiv:hep-th/9604080}.

\bibitem{KRV13}
M.~Krishnapur, B.~Rider, B.~Virag.
\newblock Universality of the stochastic Airy operator.
\newblock {\em arXiv:1306.4832}.

\bibitem{TYsys}
A.~Kuniba, T.~Nakanishi, J.~Suzuki.
\newblock $T$-systems and $Y$-systems in integrable systems.
\newblock {\em J. Phys. A}, 44:103001, 2011; {\em arXiv:1010.1344}.

\bibitem{LanTak12}
E.~Langmann, K.~Takemura.
\newblock Source identity and kernel functions for Inozemtsev-type systems.
\newblock {\em J. Math. Phys.}, 53:082105, 2012; {\em arXiv:1202.3544}.  

\bibitem{PCLevOlTak}
A.~Levin, M.~Olshanetsky.
\newblock Painlev\'e-Calogero correspondence.
\newblock {\em CRM series in Math. Phys.}, pp. 313--332, Springer, 2000; {\em arXiv:alg-geom/9706010}.  \\
K.~Takasaki.
\newblock Painlev\'e-Calogero correspondence revisited.
\newblock {\em J. Math. Phys.}, 42:1443--1473, 2001; {\em arXiv:math/0004118}.

\bibitem{LitILW}
A.~Litvinov.
\newblock On spectrum of ILW hierarchy in conformal field theory.
\newblock {\em J. High Energy Phys.}, 11:155, 2013; {\em arXiv:1307.8094}.

\bibitem{LLNZ}
A.~Litvinov, S.~Lukyanov, N.~Nekrasov, A.~Zamolodchikov.
\newblock Classical conformal blocks and Painlev\'e VI.
\newblock {\em arXiv:1309.4700v2}.

\bibitem{Mac}
I.~Macdonald.
\newblock Symmetric functions and Hall polynomials.
\newblock {\em 2nd edition, Oxford University Press}, 1995.

\bibitem{Mano12}
T.~Mano.
\newblock Determinant formula for solutions of the Garnier system and Pad\'e approximation.
\newblock {\em J. Phys. A}, 45:135206, 2012.

%\bibitem{MansHeat99}
%E.~Mansfield.
%\newblock The Nonclassical Group Analysis of the Heat Equation.
%\newblock {\em Journ. Math. Anal. Appl.}, 231:526--542, 1999.

\bibitem{MMM11}
A.~Marshakov, A.~Mironov, A.~Yu.~Morozov.
\newblock On AGT relations with surface operator insertion and a stationary limit of beta-ensembles.
\newblock {\em J. Geom. Phys.}, 61:1203--1222, 2011; {\em arXiv:1011.4491}.

\bibitem{MarTak}
K.~Maruyoshi, M.~Taki.
\newblock Deformed prepotential, quantum integrable system and Liouville field theory.
\newblock {\em Nucl. Phys. B}, 841:388--425, 2010; {\em arXiv:1006.4505}.

%\bibitem{MMShRes}
%A.~Mironov, A.~Yu.~Morozov, A.~Popolitov, Sh.~Shakirov.
%\newblock Resolvents and Seiberg-Witten representation for a Gaussian $\beta$-ensemble.
%\newblock {\em Theor. Math. Phys.}, 171:505--522, 2012; {\em arXiv:1103.5470}.

\bibitem{Nag11}
H.~Nagoya.
\newblock Hypergeometric solutions to Schr\"odinger equations for the quantum Painlev\'e equations.
\newblock {\em J. Math. Phys.}, 52:083509, 2011; {\em arXiv:1109.1645}.

%\bibitem{Nag11H}
%H.~Nagoya.
%\newblock Realizations of affine Weyl group symmetries on the quantum Painlev\'e equations by fractional calculus.
%\newblock {\em Lett. Math. Phys.}, 102:297--321, 2012.

\bibitem{NazSkl12}
M.~Nazarov, E.~Sklyanin.
\newblock Sekiguchi-Debiard operators at infinity.
\newblock {\em Comm. Math. Phys.}, 324:831--849, 2013; {\em arXiv:1212.2781v2}.

\bibitem{NazSkl13}
M.~Nazarov, E.~Sklyanin.
\newblock Integrable hierarchy of the quantum Benjamin-Ono equation.
\newblock {\em SIGMA}, 9:078, 2013; {\em arXiv:1309.6464}.

\bibitem{NekOk}
N.~Nekrasov.
\newblock Seiberg-Witten prepotential from instanton counting.
\newblock {\em Adv. Theor. Math. Phys.}, 7:831, 2004; {\em arXiv:hep-th/0206161}.  \\
N.~Nekrasov, A.~Okounkov.
\newblock Seiberg-Witten theory and random partitions.
\newblock {\em Progr. Math.}, 244:525--596, 2006; {\em arXiv:hep-th/0306238}.

\bibitem{NS09}
N.~Nekrasov, S.~Shatashvili.
\newblock Quantization of integrable systems and four-dimensional gauge theories.
\newblock {\em arXiv:0908.4052}.

\bibitem{NRS}
N.~Nekrasov, A.~Rosly, S.~Shatashvili.
\newblock Darboux coordinates, Yang-Yang functional, and gauge theory.
\newblock {\em Nucl. Phys. B (Proc. Suppl.)}, 216:69--93, 2011; {\em arXiv:1103.3919}.

\bibitem{NiSu14}
S.~Nishigaki, F.~Sugino.
\newblock Tracy-Widom distribution as instanton sum of 2D IIA superstrings.
\newblock {\em arXiv:1405.1633}.

\bibitem{DNov09}
D.~Novikov.
\newblock The $2\times2$ matrix Schlesinger system and the Belavin-Polyakov-Zamolodchikov system.
\newblock {\em Theor. Math. Phys.}, 161:1485--1496, 2009; {\em arXiv:1212.2781v2}.

%\bibitem{Ok}
%B.~Oksendal.
%\newblock Stochastic Differential Equations.
%\newblock {\em Springer, 6th edition}, 2005.

\bibitem{Piat13}
M.~Piatek.
\newblock Classical torus conformal block, $\mathcal N = 2^*$ twisted superpotential and the accessory parameter of Lam\'e equation.
\newblock {\em J. High Energy Phys.}, 1403:124, 2014; {\em arXiv:1309.7672v3}.

\bibitem{RR08}
J.~Ramirez, B.~Rider.
\newblock Diffusion at the random matrix hard edge.
\newblock {\em Comm. Math. Phys.}, 288:887--906, 2009; {\em arXiv:0803.2043v4}.

\bibitem{RRV}
J.~Ramirez, B.~Rider, B.~Virag.
\newblock Beta ensembles, stochastic Airy spectrum and diffusion.
\newblock {\em J. Amer. Math. Soc.}, 24:919--944, 2011; {\em arXiv:math/0607331}.

\bibitem{RRZ}
J.~Ramirez, B.~Rider, O.~Zeitouni.
\newblock Hard edge tail asymptotics.
\newblock {\em Elect. Comm. in Probab.}, 16:741--752, 2011; {\em arXiv:1109.4121}.

%\bibitem{RevYor}
%D.~Revuz, M.~Yor.
%\newblock Continuous Martingales and Brownian Motion.
%\newblock {\em Springer, 3rd edition}, 1999.

\bibitem{Ros13}
H.~Rosengren.
\newblock Special polynomials related to the supersymmetric eight-vertex model. II. Schr\"odinger equation.
\newblock {\em arXiv:1312.5879v2}, 2013.

\bibitem{HBP3}
I.~Rumanov.
\newblock Hard edge for beta-ensembles and Painlev\'e III.
\newblock {\em Intern. Math. Res. Notices}, rnt170, 2013; {\em arXiv:1212.5333}, 2012.

\bibitem{betaFP1}
I.~Rumanov.
\newblock Classical integrability for beta-ensembles and general Fokker-Planck equations.
\newblock {\em arXiv:1306.2117v2}, 2013.

\bibitem{betaTWk3}
I.~Rumanov.
\newblock Painlev\'e representation of Tracy-Widom$_\beta$ distribution for $\beta = 6$.
\newblock {\em to appear}, 2014.

\bibitem{Shim01}
S.~Shimomura.
\newblock Pole loci of solutions of a degenerate Garnier system.
\newblock {\em Nonlinearity}, 14:193--203, 2001.

\bibitem{SLAl93}
B.~Simons, P.~Lee, B.~Altshuler.
\newblock Matrix models, one-dimensional fermions and quantum chaos.
\newblock {\em Phys. Rev. Lett.}, 72:64--67, 1994.
\newblock Exact results for quantum chaotic systems and one-dimensional fermions from matrix models.
\newblock {\em Nucl. Phys. B}, 409[FS]:487--508, 1993.

\bibitem{Slav00}
S.~Slavyanov.
\newblock Isomonodromic deformations of Heun and Painlev\'e equations.
\newblock {\em Theor. Math. Phys.}, 123:744--753, 2000.

\bibitem{Sul9408}
B.~Suleimanov.
\newblock Hamiltonian structure of Painlev\'e equations and the method of isomonodromic deformations.
\newblock {\em Diff. Eqs.}, 30:726--732, 1994.   \\
\newblock ``Quantizations" of the second Painlev\'e equation and the problem of the equivalence of its L--A pairs.
\newblock {\em Theor. Math. Phys.}, 156:1280--1291, 2008.

\bibitem{TahZog}
L.~Takhtajan, P.~Zograf.
\newblock Hyperbolic 2-spheres with conical singularities, accessory parameters and K\"ahler metrics on $\mathcal M_{0,n}$.
\newblock {\em Trans. Amer. Math. Soc.}, 355:1857--1867, 2003; {\em arXiv:math/0112170}.

\bibitem{Tesch10}
J.~Teschner.
\newblock Quantization of the Hitchin moduli spaces, Liouville theory, and the geometric Langlands correspondence.
\newblock {\em Adv. Theor. Math. Phys.}, 15:471--564, 2011; {\em arXiv:1005.2846v5}, 2010.
  
\bibitem{TW-Airy}
C.A.~Tracy, H.~Widom.
\newblock Level-spacing distributions and the Airy kernel.
\newblock {\em Commun. Math. Phys.}, 159:151--174, 1994; {\em arXiv:hep-th/9211141}.

%\bibitem{TW-Bes}
%C.A.~Tracy, H.~Widom.
%\newblock Level spacing distributions and the Bessel kernel.
%\newblock {\em Commun. Math. Phys.}, 161:289--309, 1994.

\bibitem{TW-OrtSym}
C.A.~Tracy, H.~Widom.
\newblock On orthogonal and symplectic matrix ensembles.
\newblock {\em Commun. Math. Phys.}, 177:727--754, 1996; {\em arXiv:solv-int/9509007}.

\bibitem{Vor87}
A.~Voros.
\newblock Spectral functions, special functions and the Selberg Zeta function.
\newblock {\em Commun. Math. Phys.}, 110:439--465, 1987. \\
\newblock Zeta-regularization for exact WKB resolution of a general 1D Schr\"odinger equation.
\newblock {\em arXiv:1202.3200v2}, 2012.

\bibitem{Ya2011}
Y.~Yamada.
\newblock A quantum isomonodromy equation and its application to $\mathcal N = 2$ $SU(N)$ gauge theories.
\newblock {\em J. Phys. A}, 44:055403, 2011; {\em arXiv:1011.0292}.

%\bibitem{ZaBetheHir}
%A.~Zabrodin.
%\newblock Bethe ansatz and Hirota equation in integrable models.
%\newblock {\em arXiv:1211.4428}, 2012.

\bibitem{ZZQPC}
A.~Zabrodin, A.~Zotov.
\newblock Quantum Painlev\'e-Calogero correspondence. 
\newblock {\em J. Math. Phys.}, 53:073507, 2012; {\em arXiv:1107.5672}.
\newblock Quantum Painlev\'e-Calogero correspondence for Painlev\'e VI. 
\newblock {\em J. Math. Phys.}, 53:073508, 2012; {\em arXiv:1107.5672}.

\bibitem{ZZFP}
A.~Zabrodin, A.~Zotov.
\newblock Classical-Quantum Correspondence and Functional Relations for Painlev\'e Equations.
\newblock {\em arXiv:1212.5813}, 2012.

%\bibitem{ZenSan08}
%A.~Zenchuk, P.~Santini.
%\newblock The remarkable relations among PDEs integrable by the inverse spectral transform method, by the method of characteristics and by the Hopf-Cole transformation.
%\newblock {\em J. Phys. A}, 41:185209, 28pp, 2008.

%\bibitem{UniUE}
%I.~Rumanov.
%\newblock Universal structure and universal PDE for unitary ensembles.
%\newblock {\em Journ. Math. Phys. 51}, 083512, 2010; {\em arXiv:/0910.4417}.

%\bibitem{Multipt1matr}
%I.~Rumanov.
%\newblock All the lowest order PDE for spectral gaps of Gaussian matrices.
%\newblock {\em arXiv:/1008.3560}, 2010.

%\bibitem{TW1}
%C.A.~Tracy, H.~Widom.
%\newblock Fredholm determinants, differential equations and matrix models.
%\newblock {\em Commun. Math. Phys.}, 163:33--72, 1994.

}

\end{thebibliography}
\end{document}